\documentclass[aps,superscriptaddress,twoside,twocolumn,floatfix,a4paper,groupedaddress, pra]{revtex4-1}

\usepackage{graphicx,epsfig}
\usepackage{color}
\usepackage[usenames,dvipsnames]{xcolor}
\usepackage[ocgcolorlinks,colorlinks=true,linkcolor=blue,citecolor=red]{hyperref}
\usepackage{amsmath,amssymb, amsthm, amsfonts}
\usepackage{dsfont} 
\usepackage{xspace}
\usepackage{xcolor}
\usepackage{changes}
\usepackage{verbatim}
\usepackage{mathtools}
\usepackage{float}
\usepackage{comment}
\usepackage{soul}

\DeclareMathAlphabet{\mathbbold}{U}{bbold}{m}{n}

\newcommand{\one}{\mathbbold{1}}

\newcommand{\bb}[0]{\begin{eqnarray}}
\newcommand{\ee}[0]{\end{eqnarray}}
\newcommand{\ket}[1]{\vert #1\rangle}
\newcommand{\bra}[1]{\langle#1\vert}

\begin{document}

\title{Signatures of Liouvillian exceptional points in a quantum thermal machine}

\author{Shishir Khandelwal, Nicolas Brunner and G\'eraldine Haack}
\email{geraldine.haack@unige.ch}
\affiliation{Department of Applied Physics, University of Geneva, 1211 Gen\`eve, Switzerland}

\vspace{10pt}

\date{\today}

\begin{abstract}

Viewing a quantum thermal machine as a non-Hermitian quantum system, we characterize in full generality its analytical time-dependent dynamics by deriving the spectrum of its non-Hermitian Liouvillian for an arbitrary initial state. We show that the thermal machine features a number of Liouvillian exceptional points (EPs) for experimentally realistic parameters, in particular a third-order exceptional point that leaves signatures both in short and long-time regimes. Remarkably, we demonstrate that this EP corresponds to a regime of critical decay for the quantum thermal machine towards its steady state, bearing a striking resemblance with a critically damped harmonic oscillator. These results open up exciting possibilities for the precise dynamical control of quantum thermal machines exploiting exceptional points from non-Hermitian physics and are amenable to state-of-the-art solid-state platforms such as semiconducting and superconducting devices.

\end{abstract}

\maketitle

\section{Introduction}

A quantum thermal machine is an open quantum system coupled to thermal reservoirs. By judiciously designing the machine, one can take advantage of the energy exchange with the reservoirs in order to perform a thermodynamic task, such as cooling or producing work \cite{Kosloff2014, Vinjanampathy2016, Mitchison, Bhattha20}, or to perform a genuine quantum task, such as creating quantum correlations \cite{BohrBrask2015a, Tavakoli2018, Khandelwal2020, Heineken2020}. Formally, under a weak-coupling assumption between the system and the reservoirs, the dynamics of these machines can be modelled by master equations represented by a Liouvillian superoperator \cite{Breuer02, Rivas12, Kashuba13, Schaller14, Hofer2017}. Outside the steady-state regime, characterizing their dynamics is notably challenging \cite{Prosen08, Brask15b, Kshetrimayum17, Mitchison20, Brenes20, Haack20, Goold20}.

Open quantum systems have a natural connection to non-Hermitian physics \cite{Ganainy18, Bender07, Rotter09, Moiseyev11}. Their time evolution, captured by the Liouvillian, usually accounts for a free Hamiltonian evolution (i.e. Hermitian contribution) and for dissipation due to coupling to the reservoirs, this one being clearly non-Hermitian \cite{Minganti2018}. A hallmark of non-Hermitian physics is the possibility for the system to reach exceptional points (EPs) \cite{Heiss2012, Minganti2019, Miri19}; dissipative open quantum systems can therefore exhibit Liouvillian EPs (to be contrasted to Hamiltonian EPs that can appear in a closed quantum system described by a non-Hermitian Hamiltonian). By definition, EPs are specific points in parameter space at which two or more eigenvalues of a non-Hermitian matrix and their corresponding eigenvectors coalesce \cite{Kato}. Recently, Hamiltonian EPs have attracted lots of interest in the context of quantum sensing \cite{Am15,Chen2017,Hodaei2017,Lau18,H2019,Djorwe2019} and have been demonstrated experimentally on photonics \cite{Peng14, Gao15, Ozdemir19} and superconducting \cite{Partanen18, Naghiloo19} platforms by engineering non-Hermitian Hamiltonians.

In contrast, Liouvillian EPs have only been discussed for simple systems, such as a single dissipative spin \cite{Hatano19, Minganti2019} or in the absence of quantum jumps (i.e. considering a semi-classical approach to the dynamics) \cite{Naghiloo19}. Open questions concern the search for Hamiltonian and Liouvillian EPs in state-of-the-art physical platforms and their signatures \cite{Dietz07, Cartarius11, Demange12,Uzdin2013,Gilary13, Garmon17, Fernando18,Kosloff2017, Insinga18, Chakraborty19, Cabot19, Archak20}, especially in the quantum regime.

\begin{figure}[t]
\includegraphics[width=0.9\columnwidth]{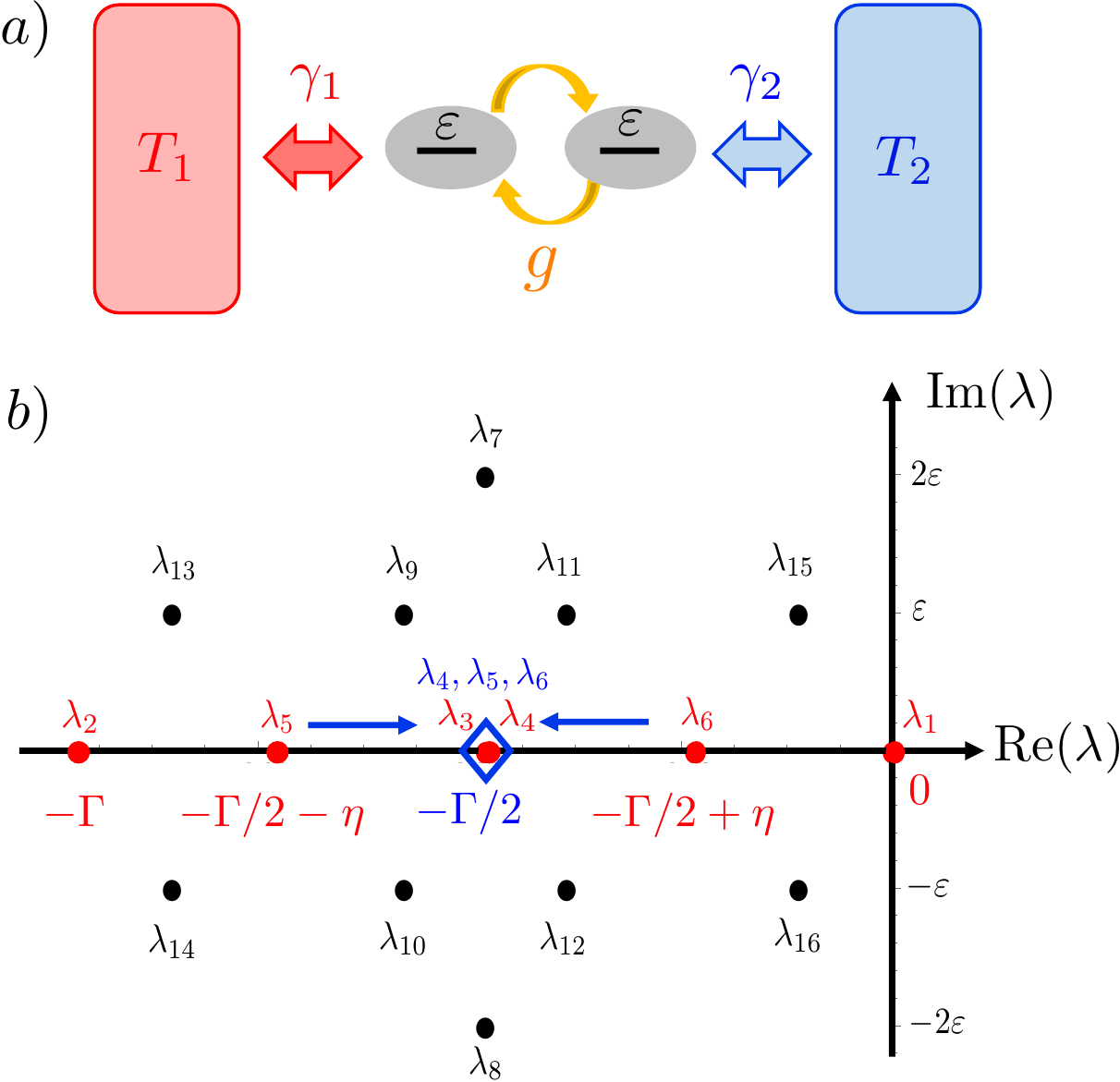}
\caption{a) Model of a two-qubit entanglement engine. Parameters of the model are the bare energies $\varepsilon$ of the two qubits, the inter-qubit coupling strength $g$, the temperatures of the two baths $T_{1,2}$, and the couplings to the baths $\gamma_{1,2}$.b) Spectrum $\sigma(\mathcal{L})$ of the full Liouvillian (in black) and spectrum of the reduced Liouvillan $\sigma(\tilde{\mathcal{L}})$ (in red) showing the coalescence of $\lambda_4$, $\lambda_5$ and $\lambda_6$ at a third-order EP (in blue). $\Gamma$ and $\eta$ depend on the parameters of the model, see main text.
}
\label{fig:machine}
\end{figure}

In this work, we solve analytically the dynamics of a quantum thermal machine, a non-Hermitian quantum system. We demonstrate the existence of Liouvillian EPs for experimentally accessible range of parameters. We uncover a signature of EPs in the long-time dynamics, in the form of critical decay towards the steady state, in analogy to critical damping in an harmonic oscillator. These results broaden the class of systems exhibiting EPs and opens new routes for controlling the quantum dynamics of nanoscale thermal machines.

The paper is organized as follows. In Sec. \ref{sec:model} we describe the model for a two-qubit entanglement engine and provide its exact dynamics in the transient regime by solving the corresponding Lindblad master equation and deriving the spectrum of the non-Hermitian Liouvillian. The solution is provided in terms of a unique decomposition of the density operator for the two-qubit engine in terms of the eigenvalues and eigenvectors of the Liouvillian. In Sec. \ref{sec:LEP}, we demonstrate the existence of EPs analytically. In Sec. \ref{sec:short}, we provide signatures of EPs in the short-time dynamics. These signatures, associated with polynomial-in-time dependance, are in agreement with previous works on signatures of Hamiltonian and Liouvillian EPs \cite{Cartarius11}. In Sec. \ref{sec:long}, we put forward a novel striking consequence of EP on the dynamics of the two-qubit entanglement engine. We demonstrate critical damping of the entanglement engine towards its steady state at EP and discuss its consequence on the heat current and entanglement generation. Results in Secs. \ref{sec:short} and \ref{sec:long} are discussed considering set of parameters for state-of-the-art experiments with solid-state platforms, see Ref.~\cite{BohrBrask2015a} for more references. We conclude in Sec. \ref{sec:conc}.


\section{Model and exact transient dynamics}\label{sec:model}

We consider the entanglement engine shown in Fig.~\ref{fig:machine} that consists of two interacting qubits, each of them incoherently and weakly coupled to its own thermal reservoir. The machine is autonomous in the sense that it operates without an external source of coherence or control. In the presence of a thermal gradient between the reservoirs, it was shown that the heat current through the machine sustains steady-state entanglement \cite{BohrBrask2015a, Khandelwal2020}. Under a weak system-bath interaction assumption, the dynamics of this open quantum system is described by a Lindblad master equation \cite{Breuer02, BohrBrask2015a}. Its specific form depends in particular on the relative strength of the inter-qubit interaction strength with respect to the system-bath coupling strength, leading to the so-called local or global master equations \cite{Hofer2017, Khandelwal2020}. The master equation can be expressed in the form, $\dot\rho\left(t\right) = \mathcal{L} \rho(t)$, $\mathcal{L}$ being a Liouvillian superoperator acting on the $4\times4$ reduced density operator $\rho$ of the two qubits.  For our analysis, it is essential to write $\mathcal{L}$ in a $16\times16$ matrix-operator representation, $L$  acting on the vectorized form $\boldsymbol  p$ of $\rho$. We refer to App. \ref{app:1a} for the explicit form of the master equation and of $L$.



The Liouvillian being non-Hermitian, its eigenvalue spectrum $\sigma(\mathcal{L})$ is complex and there exist distinct right and left eigenmatrices, $\rho_i$ and $\sigma_i$, defined by $\mathcal L\rho_i = \lambda_i\rho_i$ and $\sigma_i^{\dagger}\mathcal L = \lambda_i\sigma_i^{\dagger}$ respectively \cite{Macieszczak2016,Minganti2018,Minganti2019}. They are bi-orthogonal with respect to the Hilbert-Schmidt inner product, $\langle\sigma_i,\rho_j\rangle \coloneqq \text{Tr}\left(\sigma_i^{\dagger}\rho_j\right)= \delta_{ij}$. The full spectrum $\sigma\left(\mathcal{L}\right)$ is shown in Fig.~\ref{fig:machine}b). The points in parameter space where $n$ eigenvalues and the corresponding eigenmatrices coalesce correspond to $n\text{th}$-order EPs. A full analytical characterization of EPs of a non-Hermitian matrix depends on feasibility of eigenvalue computation, which in turn depends on the degree to which the characteristic polynomial of the matrix can be factorized. For the two-qubit entanglement engine, we conduct this analysis completely and find an array of second, third and fourth-order EPs \cite{Ding16, Hatano19, Bhattacherjee19, Dey20}. In App. \ref{app:full_spectrum}, we provide the analytical expressions of the 16 eigenvalues of the spectrum in terms of the parameters of the model.

For further analysis, leading up to signatures of EPs, we reduce the dimension of the problem by exploiting the form of $H_{\text{int}}$. The latter imposes the steady state of the two qubits to only have six possible non-zero elements (the populations of the four computational states and the two coherences, $\ket{10}\bra{01}$ and $\ket{01}\bra{10}$), all other elements decaying exponentially to zero. Hence, we consider initial states that belong to the steady-state subspace, which reduces the $16\times16$ Liouvillian $L$ to a $6\times6$ Liouvillian $\tilde{L}$, whose exact form is provided in App. \ref{app:compact}. 

Its spectrum $\sigma(\tilde{\mathcal{L}})$ is highlighted in red in Fig.~\ref{fig:machine}~b), and reduces to 
\bb
\label{eq:spectrum}
\sigma(\tilde{\mathcal{L}}) &=& \{ \lambda_i\}\,\, i=1,\ldots, 6 \nonumber \\
&=& \{0;- \Gamma; -\frac{\Gamma}{2}; -\frac{\Gamma}{2}; -\frac{\Gamma}{2} - \eta; -\frac{\Gamma}{2} + \eta   \}
\ee
with 
\bb
\label{eq:eta}
\eta \coloneqq \sqrt{\Delta \Gamma^2 - 4g^2}\,.
\ee
Here we have introduced the simplified notation $\Gamma_j\coloneqq \gamma_j^-+\gamma_j^+$, $\Gamma\coloneqq \Gamma_1 + \Gamma_2$ and $\Delta \Gamma \coloneqq (\Gamma_1-\Gamma_2)/2$. At all points in parameter space that are not exceptional points, $L$ is diagonalizable with eigenvalues $\lambda_i ,\, i=1,\ldots,6$, such that $\text{Re}\left(\lambda_i\right)\leq0$ \cite{Rivas12}. Furthermore, we have a unique steady state which corresponds to the eigenvalue 0. The general solution of the master equation, $\rho(t) = e^{\mathcal{L} t} \rho(0)$, can be written as a weighted sum of the right eigenmatrices $\rho_i$ with exponentially decaying factors \cite{Macieszczak2016,Minganti2019}
\bb
\label{eq:deco}
\rho(t, \xi) = \rho_{\text{ss}}(\xi) + \sum_{i=2}^{6} c_i(\xi) \, e^{\lambda_i(\xi) t}  \rho_i(\xi)\,,
\ee
with  $ \rho_{\text{ss}} = \rho_1$, the steady state corresponding to the zero eigenvalue ($\lambda_1 = 0$), and $\sigma_1 = \one$  being the identity matrix by construction. The dynamics depends on all parameters of the model, which we denote by the set $\xi = \{\varepsilon, T_1, T_2, \gamma_1, \gamma_2,g\}$ and on the the initial state through the coefficients $c_i(\xi)\coloneqq \left\langle \sigma_i\left(\xi\right),\rho\left(0\right)\right\rangle$, corresponding to its overlap with the left eigenmatrices. Note that by construction, right eigenmatrices $\rho_i$ are Hermitian and normalized with the Hilbert-Schmidt norm \cite{Minganti2018,Minganti2019}. Under this convention, the decomposition Eq.~\eqref{eq:deco} is unique.\\


\section{Liouvillian exceptional points}\label{sec:LEP}

This model for an entanglement engine features a remarkable variety of Liouvillian EPs. When considering the full Liouvillian matrix $L$, we demonstrate the mathematical existence of second-, third-, and fourth-order Liouvillian EPs. The full spectrum and list of EPs is provided in App. \ref{app:full_spectrum} and \ref{app:full_EP}. We also considered a weak and strong inter-qubit interaction regime (see App. \ref{app:strong}), and showed that different EPs exist in these two regimes when considering the full Liouvillian that includes decay modes for off-diagonal elements vanishing in the steady-state (to be contrasted to recent results in \cite{Scali20}).

When considering the reduced Liouvillian matrix $\tilde{L}$ that dictates the dynamics of non-zero elements in the steady-state, the situation is different. Under strong inter-qubit interaction strength, no EP exist. In contrast, under a weak coupling assumption, we demonstrate the existence of a third-order EP. From Eqs.~\eqref{eq:spectrum} and \eqref{eq:eta}, it is straightforward to see that, when $\eta$ becomes 0, the eigenvalues $\lambda_5$ and $\lambda_6$ merge with $\lambda_3$ and $\lambda_4$. However, only the eigenmatrices of $\lambda_4, \lambda_5$ and $\lambda_6$ coalesce as shown in App. \ref{app:eigenmatrices}, leading to a third-order exceptional point, with the merged eigenvalue $\bar{\lambda} = -\Gamma/2$, see the blue diamond in Fig.~\ref{fig:machine}b). 
To reach this EP, $\eta=0$, we choose to only vary the inter-qubit coupling $g$ while keeping constant the couplings to the baths entering the definition of $\Delta \Gamma$. We denote by 
$\bar{g}$ the solution for $g$ satisfying $\eta = 0$, and the corresponding set of parameters by $\bar{\xi}$. \\

At an EP, the geometric multiplicity of the merged eigenvalue is less than its algebraic multiplicity. This necessitates constructing generalised eigenmatrices to obtain a complete basis for the Liouvillian at the EP. At $\eta=0$ (i.e. $g = \bar{g}$, also $\xi = \bar{\xi}$), a third-order EP is reached, $\lambda_4(\bar{\xi}) = \lambda_5(\bar{\xi}) = \lambda_6(\bar{\xi}) \equiv \bar{\lambda}$ and we need to construct two generalised eigenmatrices, $\rho^\prime$ and $\rho^{\prime\prime}$, such that the subspace corresponding to the third-order EP will be spanned by $\{ \rho_4, \rho^\prime, \rho^{\prime\prime} \}$. The generalized eigenmatrices can be determined by solving a Jordan chain \cite{Horn2012}, given by $({\mathcal{L}}- \bar{\lambda})\bar\rho_4 = 0$, $({\mathcal{L}}- \bar{\lambda}) \rho^\prime = \alpha \bar\rho_4$, $({\mathcal{L}}-\bar{\lambda}) \rho^{\prime\prime}= \beta \rho^{\prime}$ and similarly for left eigenmatrices. The parameters $\alpha$ and $\beta$ are complex coefficients depending on the parameters $\bar{\xi}$, ensuring normalization of the generalized right eigenvectors, see App. \ref{app:gen}. At the EP $\eta=0$, the unique spectral decomposition Eq.~\eqref{eq:deco} is then transformed into:
\bb
\label{eq:dyn_EP}
&&\bar\rho(t,\bar{\xi}) = \bar\rho_{\text{ss}}(\bar{\xi}) +  \sum_{i=2,3} \bar{c}_i(\bar{\xi}) e^{\lambda_i (\bar{\xi}) t} \bar\rho_i(\bar{\xi}) \nonumber \\
 && + \left( \bar{c}_4(\bar{\xi}) +\alpha(\bar{\xi}) \,  \bar{c}_5(\bar{\xi})  t +\alpha(\bar{\xi})\, \beta(\bar{\xi}) \,   \bar{c}_6(\bar{\xi})  \frac{t^2}{2} \right) e^{\bar{\lambda} t} \bar\rho_4(\bar{\xi}) \nonumber \\
&&+ \Big( \bar{c}_5(\bar{\xi})  + \beta(\bar{\xi}) \,  \bar{c}_6(\bar{\xi})  t \Big) e^{\bar{\lambda} t} \rho^{\prime}(\bar{\xi}) +  \bar{c}_6(\bar{\xi})  e^{\bar{\lambda} t} \rho^{\prime\prime}(\bar{\xi})\,.
\ee
Here we recall $\bar{\xi} \coloneqq \{\varepsilon, T_1, T_2, \gamma_1, \gamma_2, \bar{g}\}$ with $\bar{g}$ such that $\eta=0$. The coefficients $\bar{c}_i(\bar{\xi}) \coloneqq  \left\langle \bar{\sigma}_i\left(\bar{\xi}\right), \rho\left(0\right) \right\rangle$ depend on the left generalized eigenvectors $\bar{\sigma}_i$, which are constructed bi-orthogonal with the generalized right eigenmatrices. Mathematically, this unique decomposition shows a characteristic time-behaviour at EP, with time-polynomial factors entering the transient dynamics in the subspace corresponding to the EP, spanned by the generalized eigenmatrices, $\rho_4, \rho^\prime$ and $\rho^{\prime\prime}$.

\begin{figure}[t!]
	\centering
	\hspace*{-0.3cm}\includegraphics[width=1\columnwidth]{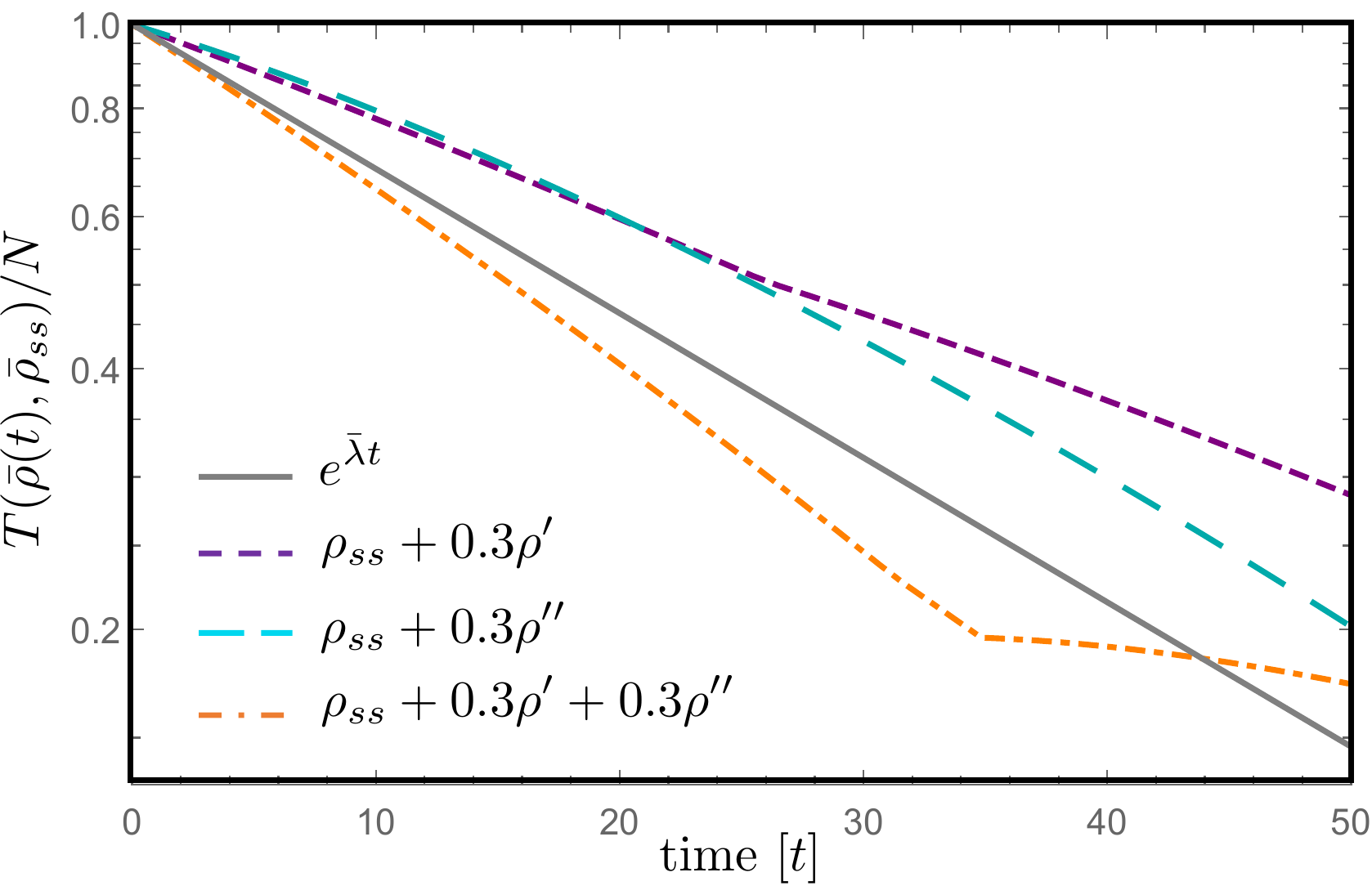}\caption{Log-plot of the trace distance normalized by its maximum value $N := T(\bar{\rho}(t=0), \bar{\rho}_{\text{ss}})$ as a function of time, for 3 different initial states that evidence the departure from a purely exponentially decaying behaviour (gray solid line) at short times at the EP. The three initial states are constructed as a weighted sum of the steady-state and the right generalized eigenvectors $\rho'$ and $\rho''$, constituting a set of states of measure 0. Set of parameters $\xi=\{\varepsilon=1,T_1=3,T_2=0.7, \gamma_1 = \gamma_2 = 0.01, \bar{g}\}$. 
	}
	\label{fig:dyn}
\end{figure}

\section{Signatures in the short-time dynamics}\label{sec:short}

The effect of the time-polynomial terms in the decomposition Eq.~\eqref{eq:dyn_EP} strongly depends on the initial state that will set the coefficients $\bar{c}_i(\bar{\xi})$ and will in general be difficult to distinguish from a sum of purely exponential decay, see Eq.~\eqref{eq:deco}. To evidence this time-polynomial behaviour, a possibility is to construct mathematically an initial state 
such that its time-evolution operates exclusively in the subspace of the EP to evidence the Jordan block structure at the EP. This mathematical construction amounts to a weighted sum of the steady state $\rho_{\text{ss}}$ and the eigenmatrices $\rho_5$ and $\rho_6$ at non-EP and of $\rho'$ and $\rho''$ at the EP, ensuring the corresponding density operator to be Hermitian and positive semi-definite.  In Fig.~\ref{fig:dyn},  we consider three different weighted sums of $\rho_{\text{ss}}, \rho'$ and $\rho''$ as initial states that probe the transient dynamics through the decay modes affected by the EP. A clear departure from a purely exponential decay set by $\bar{\lambda}$ (gray solid curve) at EP is evidenced (shown on a logarithmic scale). This short-time behaviour is known and is in agreement with previous works on EPs and signatures at short times, in particular in the context of survival probability \cite{Cartarius11}. The decay modes with a time-polynomial dependence are evidently suppressed at long-times.\\


\section{Critical damping}\label{sec:long}

\begin{figure}[t!]
	\centering
	\includegraphics[width=\columnwidth]{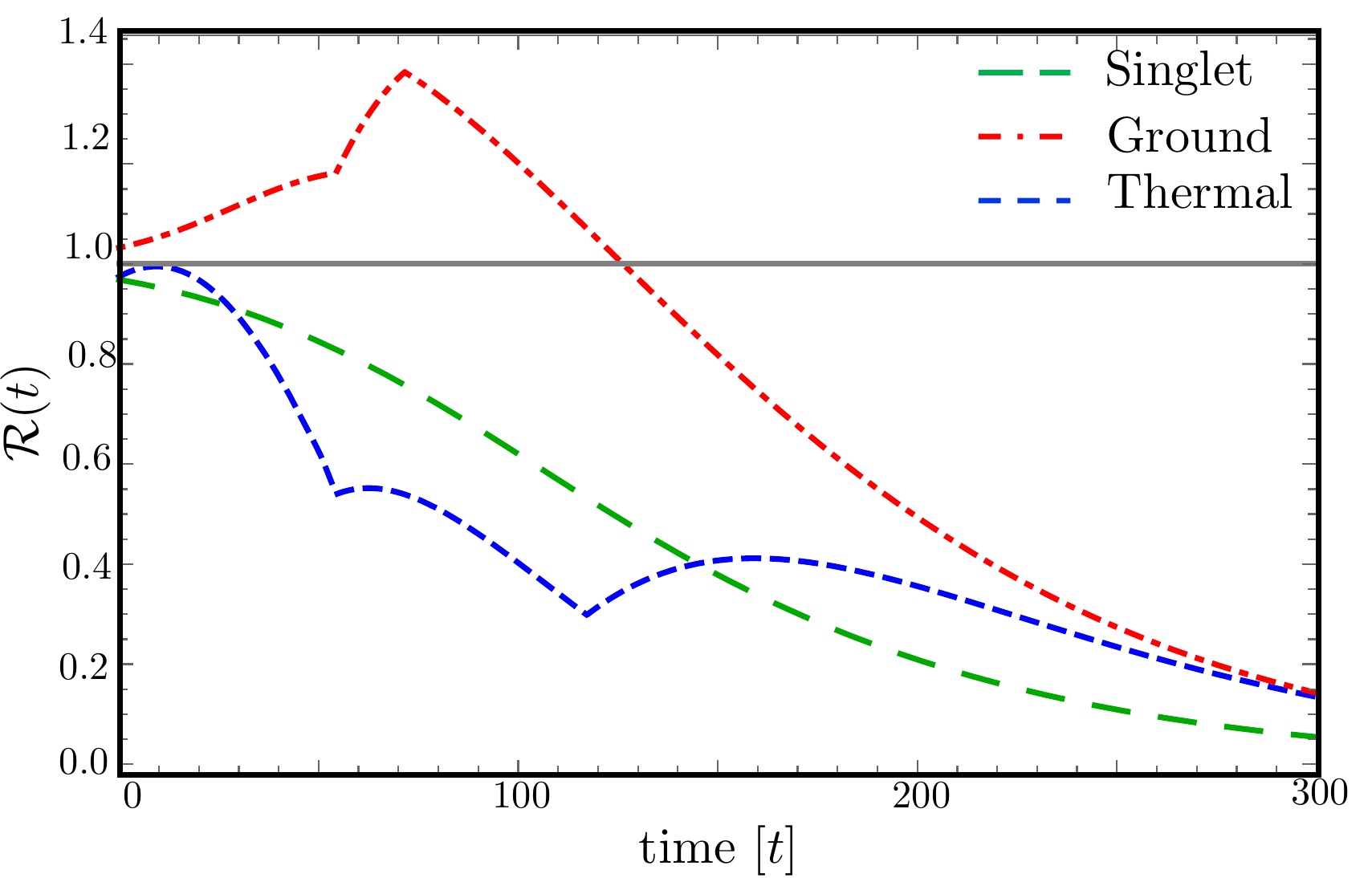}\caption{Critical damping at EP evidenced by the ratio $\mathcal{R}(t)$ (see Eq.~\eqref{eq:long_ratio}) for a variety of initial states: qubits in a singlet state (green), in their ground state (red) and in a product of thermal states set by the temperatures $T_1$ and $T_2$ (blue). $\mathcal{R} = 1$ (black, dashed) implies that the trace distance to the steady-state is the same at EP and away from EP. Around $t \sim 130$, $\mathcal{R} < 1$ for all three cases, i.e. the system at the EP is closer to its steady state compared to the system at non-EP. The precise time at which $\mathcal{R}$ becomes smaller than 1 depends on the initial state. At very long times, $\mathcal{R} \rightarrow 0$ for all three states. Set of parameters $\xi = \{\varepsilon = 1, T_1=3, T_2 = 0.7, \gamma_1 = \gamma_2 = 0.01, g = 0.005\}$ at the non-EP. At the EP, $g \rightarrow \bar{g}\approx 0.011$ to satisfy $\eta=0$.}. \label{fig:critical}
\end{figure}

Whereas polynomial-time dependence is expected from Eq.~\eqref{eq:dyn_EP} and has been discussed in the literature, in this section, we evidence a novel signature to be identified in the long-time dynamics of an open quantum system at the EP. To this end, we investigate how the state of the quantum thermal machine relaxes towards its steady state, \textit{i.e.} we characterize the dynamics through the presence or not of oscillations and determine the damping rates associated with the different decay channels. As a measure, we consider the trace distance between $\rho(t)$ and the steady state $\rho_{\text{ss}}$, $T\left(\rho(t),\rho_{\text{ss}} \right)\coloneqq\lvert\lvert \rho\left(t\right)-\rho_{\text{ss}}\rvert\rvert_1/2$ ($\lvert\lvert\cdot\rvert\rvert_1$ being the 1-norm). We compare the trace distance at the EP (for the set of parameters $\bar{\xi}$) and away from EP (for $\xi$) through the ratio: 
\bb
\label{eq:ratio}
\mathcal{R}(t) = \frac{T\left( \bar\rho(t, \bar{\xi}), \bar\rho_{\text{ss}}(\bar{\xi}) \right) }{T\left( \rho(t,\xi), \rho_{\text{ss}}(\xi) \right) } \,.
\ee
This ratio can be smaller or larger than 1 depending on how the state at the EP decays to the steady state compared to the state at a non-EP. At long times, we demonstrate that $\mathcal{R}(t) \rightarrow 0$ for all initial states (except for a very specific set of initial states of measure 0, see discussion below).
The proof is available in App. \ref{app:crit}. Alternatively, for this model, an intuition about its behaviour can easily be built by estimating it in the subspace corresponding to the EP solely:
\bb
\label{eq:long_ratio}
\mathcal{R}(t) &\sim& \frac{\left\lvert\left\lvert \left(\bar{c}_4+\alpha \bar{c}_5 t +\alpha\beta \bar{c}_6\frac{t^2}{2}\right) \rho_4
 	+\left(\bar{c}_5 + \beta \bar{c}_6t \right)\rho^{\prime} + \bar{c}_6 \rho^{\prime\prime} \right\rvert\right\rvert_1}{\left\lvert\left\lvert  c_4 \rho_4 + c_5e^{-\eta t} \rho_5 + c_6 e^{\eta t} \rho_6 \right\rvert\right\rvert_1} \nonumber \\
	&&
\ee
Equation \eqref{eq:long_ratio} is obtained by inserting the exact form of the eigenvalues $\lambda_5$ and $\lambda_6$ at EP and at a non-EP. The spectrum of the Liouvillian being symmetric with respect to the imaginary axis and the eigenvalues coming in pairs, the coalescing eigenvalues can always be expressed in terms of the EP plus or minus a real part. In our model, this takes the explicit form $\lambda_{5,6} = \bar{\lambda} \mp \eta$ and leads to the expression of the denominator at non-EP, where we have omitted the labels $\xi, \bar{\xi}$ for clarity. \par 
 Equation \eqref{eq:long_ratio} highlights three regimes for the dynamics as a function of the parameter $\eta$. If $\eta>0$, the dynamics does not exhibit oscillation, and the system at EP approaches its steady state faster than at a non-EP. If $\eta$ is imaginary, oscillations are present as the system decays towards its steady state. The specific case $\eta=0$ corresponds to a critical regime marking the boundary between the two previous cases. The critical regime corresponds to the \textit{fastest aperiodic decay}. In particular, the dynamics in the critical regime is faster than in any \textit{overdamped} situation, while preventing oscillations present in the \textit{underdamped} regime. The system therefore exhibits a remarkable similarity with a classical damped Harmonic Oscillator (HO). \\
 
 \begin{figure*}
	\centering
	\includegraphics[width=0.95\textwidth]{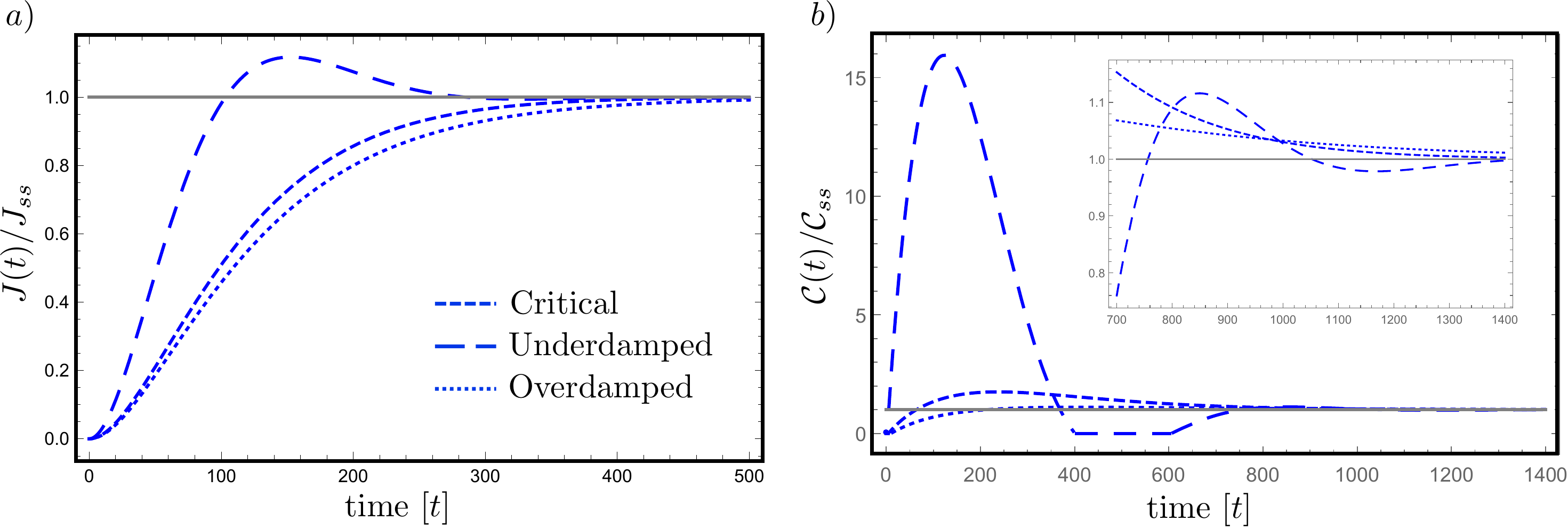}\caption{Entanglement engine operated at the EP (critical regime), compared to underdamped and overdamped regimes. Panel a) Heat current and Panel b) concurrence, both normalized by their steady-state value and plotted as a function of time, for $\eta$ imaginary ($g = 0.005$, underdamped regime, long-dashed), $\eta =0$ ($g=\bar{g}$, critical damping at the EP, dashed) and $\eta >0$ ($g = 0.001$, overdamped regime, dotted). For both panels, the two qubits are initially in a product of thermal state. Critical damping affects the dynamics of the entanglement engine, both the heat current and the concurrence show aperiodic fastest decay towards their steady-state values. Inset in Panel b) highlights long-time oscillations in the underdamped regime, as compared to overdamped and critical regimes. Set of parameters: $\xi = \{\varepsilon = 1, T_1=1, T_2 = 0.1, \gamma_1 = 0.001, \gamma_2 = 0.011\}$.}\label{fig:heat_conc}
\end{figure*}

A classical damped HO is described by the equation of motion $m\ddot x + \gamma \dot x +kx=0$, with mass $m$, spring constant $k$ and damping coefficient $\gamma$ . The damping ratio $\zeta = \gamma/(2\sqrt{km})$ determines the behaviour of the HO; overdamped for $\zeta>1$, critically damped for $\zeta=1$ and underdamped for $\zeta<1$. By expressing the equation of motion as a matrix differential equation in phase space, it can be shown that $\zeta=1$ is an exceptional point of the damped HO (see for example, \cite{Fernandez2018} and App. \ref{app:ho} for more details). For the two-qubit machine, rewriting $\eta$ as $ \eta= 2g \sqrt{\tilde{\zeta}^2-1}$ with $\tilde{\zeta} = \Delta \Gamma/(2g)$, the two-qubit machine also shows these regimes for approaching the steady state, determined by $\tilde\zeta$. The case $\tilde{\zeta} >1$ corresponds to the overdamped regime, $\tilde{\zeta} < 1$ corresponds to the underdamped regime exhibiting oscillations, and $\tilde{\zeta} = 1$ corresponds to the boundary between the two, or critical decay. In a damped HO, one may approach any of the regimes by varying either the damping coefficient or the spring constant. For the two-qubit machine, however, we vary the inter-qubit coupling, $g$. The reason for this choice is two-fold; first, upon varying $g$, the common part of the eigenvalue $\bar \lambda$ remains unchanged, and second, this method allows one control the dynamics of the machine without altering the coupling to the baths. A mathematical difference between damped regime of the classical HO and our two-qubit thermal machine lies in the nature of the EP. Our model exhibits a third-order EP, while the damped HO has a second-order EP. However, both the EPs have a square-root structure \cite{Heiss2012} and not a cube-root one as present for a single two-level system coupled to an environment \cite{Hatano19}. An important property of a damped HO is that it exhibits fastest aperiodic decay at the point of critical damping; meaning that the oscillator's decay is faster at critical damping than in any overdamped situation. This property is in fact rooted in the time-polynomial terms appearing in the dynamics at the EP.

One can realize that the dynamics at critical decay is faster than the one in the overdamped regime by considering Eq. \eqref{eq:long_ratio} at sufficiently long times for $\eta > 0$. The term proportional to $e^{-\eta t}$ in the denominator vanishes, regardless of the exact value of $c_5$. The numerator therefore grows as $\mathcal O\left(t^2\right)$, while the denominator grows as $\mathcal O\left(e^{\eta t}\right)$. The time scale of the decay is the same at critical decay and in the case where $\eta$ is imaginary (underdamped regime), however the system then undergoes oscillations indefinitely. It should be noted that critical damping towards the steady-state in the long-time dynamics is present if and only if the initial state has a finite overlap with the sixth eigenmatrix, $\bar{c}_6(\bar{\xi}) = \text{Tr} \left( \bar{\sigma}^\dagger_6(\bar{\xi}) \rho(0) \right) \neq 0\,$, which is a condition fulfilled by all initial states, except a set of states of measure 0 that would need to be constructed mathematically to not fulfil this condition. \\

Therefore, we will have 
\bb
\label{eq:ratio}
\mathcal{R} (t) <1 \quad \text{for large $t$}\,.
\ee
At the EP, the system is always closer to its steady state compared to the system evolving at non-EP with $\eta>0$, while being aperiodic (in contrast to the regime characterized by an imaginary $\eta$). The EP can therefore be seen as the point of critical damping of the two-qubit machine. The full proof is provided in App. \ref{app:crit}. Figure~\ref{fig:critical} illustrates this critical behaviour, considering different initial states for the two qubits; a product state of two thermal states at temperatures $T_1, T_2$ (blue), the two qubits in their ground state (red) and a singlet state (green). The presence of coherences (in the singlet state) does not affect the above critical time-behaviour at EP at long times.
At sufficiently long times, $\mathcal R\left(t\right)$ is clearly smaller than 1 and tends to 0 exponentially, regardless of the initial state. At short and intermediate times, the non-monotonic behaviour of $\mathcal{R}$ originates from the dependence on the initial state through the coefficient $\bar{c}_i$ and the polynomial dependence in time superimposed to the exponential decay, see Eq.~\eqref{eq:dyn_EP}. The initial state also determines the time at which the dynamics at EP will be closer the steady state compared to the dynamics at a non-EP, i.e when $\mathcal{R} <1$.

In Fig.~\ref{fig:heat_conc}, we investigate how critical damping at the EP $\eta=0$ affects the heat current flowing through the machine and the amount of entanglement present in the transient regime. 
As initial state for the two qubits, we consider for this figure the two qubits to be in a product of thermal states determined by the temperatures of their respective bath. The critical damping demonstrated for the dynamics of the density operator is also evidenced in the heat current and in entanglement generation. Both heat current and concurrence show a faster decay towards their steady-state values as compared to the overdamped regime, while being aperiodic in contrast to their dynamical behaviour in the underdamped regime. The inset highlights oscillations in the underdamped regime at long times. Critical damping corresponds to the fastest aperiodic decay. The value of $g$ that achieves critical damping is the largest one that can lead to non-oscillatory behaviour. Remarkably, a set of experimentally valid parameters can be found to reach critical damping and a finite measure for entanglement, different from the set of parameters used in the previous figures. This also illustrates the existence of multiple valid sets of parameters that can exhibit EPs in this model. These results open the way to a new type of control of the quantum dynamics of quantum thermal machines, where a suitable choice of parameters allows the entanglement engine to properly work and to optimize its dynamics towards its steady state, without any type of additional external control.\\

\section{Conclusion} \label{sec:conc}

In this work, we solved exactly the dynamics of an autonomous quantum thermal machine in the transient regime. Focusing on the non-Hermitian physics induced by the dissipators to the environments, we derived a number of Liouvillian EPs of different orders for this system for experimentally valid range of parameters. These EPs feature signatures in both short- and long-time dynamics. We derived the existence of a third-order EP when investigating specifically the decay modes controlling the non-zero elements in the steady state.  Remarkably, we demonstrated that this EP corresponds to the point of critical decay of the machine towards its steady state, in close analogy with a classical damped HO, and showed its consequence on heat current and entanglement production. Critical decay at EP for the two-qubit entanglement engine corresponds to the fastest aperiodic decay towards its steady state. 
These results open exciting new routes towards the full dynamical control of quantum thermal machines and their approach towards the steady state without additional time-dependent external drives. They also open new questions for future analysis, concerning the role of different nth-order EP onto the dynamics and their possible applications towards dynamical quantum metrology and quantum sensing. \\

\emph{Note added.} In a related work, Kumar et al. \cite{Kumar2021} discuss optimized steering of a one-qubit state using an EP. EPs have also been proposed to enhance the performance of thermal motors \cite{Fernandez2020}.

\section*{Acknowledgements}
G. H. thanks K. Murch, N. Hatano, F. Minganti for initial stimulating discussions. The authors thank the referees for constructive reports. The authors acknowledge support from the Swiss National Science Foundation through the starting grant PRIMA PR00P2\_179748 and through the NCCRs QSIT and SwissMAP.

\begin{widetext}

\appendix

\section{Two-qubit thermal machine}
\label{app:full_weak}

\subsection{Full Liouvillian}
\label{app:1a}
We consider a two-qubit thermal machine shown in Fig.~1 of the main text, consisting of two qubits interacting with strength $g$, each coupled weakly to a distinct Markovian thermal reservoir $k$ with strength $\gamma_k$. Assuming weak inter-qubit interaction such that $g\lesssim\gamma_k\ll \varepsilon$, the dynamics of the two qubits are governed by a local  Lindblad master equation \cite{Breuer02, BohrBrask2015a, Hofer2017, Khandelwal2020} of the form (we set $\hbar = k_{B}=1$ everywhere)

\begin{equation}
\label{eq:lind}
\begin{aligned}
	\dot\rho\left(t\right) = &\mathcal{L} \rho(t)
	=  -i\left[H_{\text S} +H_{\text{int}},\rho\left(t\right)\right]\\ &+ \sum_{k=1,2}\gamma_k^+\mathcal D[\sigma^{(k)}_+] \rho(t)+\gamma_k^-\mathcal D[\sigma^{(k)}_-] \rho(t)\,,
\end{aligned}
\end{equation}
where $\mathcal{L}$ is the Liouvillian superoperator, $H_{\text S} =\sum_k \varepsilon \sigma^{\left(k\right)}_+\sigma^{\left(k\right)}_-$ and $H_{\text{int}}=g\left(\sigma^{\left(1\right)}_+\sigma^{\left(2\right)}_- +\sigma^{\left(1\right)}_-\sigma^{\left(2\right)}_+\right)$. The matrices $\sigma^{\left(k\right)}_\pm$ are the raising and lowering operators corresponding to qubit $k$. Dissipation into the reservoirs is captured through the superoperators $\mathcal{D}[A]\cdot = A \cdot A^\dagger - \{ A^\dagger A, \cdot \}/2$ and the incoming and outgoing rates, $\gamma_k^+$ and $\gamma_k^-$ respectively, depend on the statistics of the reservoir, labelled $F$ for fermionic and $B$ for bosonic, $\gamma_{k,F/B}^+ = \gamma_k n_{F/B}$ and $\gamma_{k,F/B}^- = \gamma_k (1 \mp n_{F/B})$, with the minus sign for fermions and the plus sign for bosons. To characterize the evolution of the two-qubit thermal machine in the form of a non-Hermitian Liouvillian, we recast the above master equation as a matrix differential equation for the vectorized state $\boldsymbol{p}(t)$ of the density operator $\rho(t)$. 
\begin{align}\label{eq:vec}
	\rho(t)\longleftrightarrow \boldsymbol{p}(t), \quad \quad \dot{\rho}(t) = \mathcal{L}\rho(t)\longleftrightarrow \dot{\boldsymbol{p}}(t) = L\boldsymbol{p}(t), 
\end{align}where $\boldsymbol p = \left(p_1,...,p_{16} \right)^T$  with $p_1,...,p_{16}$ being the elements of $\rho(t)$ and $L$ is a $16\times 16$ matrix. Explicitly, $L$ is the matrix 
\begin{equation}
	\hspace{-0.9cm}
	\resizebox{1.1\hsize}{!}{$  \left(
		\begin{array}{cccccccccccccccc}
			-\gamma _{1}^- -\gamma _{2}^- & 0 & 0 & 0 & 0 & \gamma _{2}^+ & 0 & 0 & 0 & 0 & \gamma _{1}^+ & 0 & 0 & 0 & 0 & 0 \\
			0 & -i \varepsilon -\gamma _{1}^- - \frac{\Gamma _2}{2} & i g & 0 & 0 & 0 & 0 & 0 & 0 & 0 & 0 & \gamma _{1}^+ & 0 & 0 & 0 & 0 \\
			0 & i g & -i \varepsilon -\gamma _{2}^- -\frac{\Gamma _1}{2} & 0 & 0 & 0 & 0 & \gamma _{2}^+ & 0 & 0 & 0 & 0 & 0 & 0 & 0 & 0 \\
			0 & 0 & 0 & -\frac{1}{2} i (4 \varepsilon -i \Gamma ) & 0 & 0 & 0 & 0 & 0 & 0 & 0 & 0 & 0 & 0 & 0 & 0 \\
			0 & 0 & 0 & 0 & i \varepsilon -\gamma _{1}^- -\frac{\Gamma _2}{2} & 0 & 0 & 0 & -i g & 0 & 0 & 0 & 0 & 0 & \gamma _{1}^+ & 0 \\
			\gamma _{2}^- & 0 & 0 & 0 & 0 & -\gamma _{1}^- -\gamma _{2}^+ & i g & 0 & 0 & -i g & 0 & 0 & 0 & 0 & 0 & \gamma _{1}^+ \\
			0 & 0 & 0 & 0 & 0 & i g & -\frac{\Gamma }{2} & 0 & 0 & 0 & -i g & 0 & 0 & 0 & 0 & 0 \\
			0 & 0 & \gamma _{2}^- & 0 & 0 & 0 & 0 & -i \varepsilon -\gamma _{2}^{+}-\frac{\Gamma _1}{2} & 0 & 0 & 0 & -i g & 0 & 0 & 0 & 0 \\
			0 & 0 & 0 & 0 & -i g & 0 & 0 & 0 & \frac{1}{2} \left(2 i \varepsilon -2 \gamma _{2}^{-} -\Gamma _1\right) & 0 & 0 & 0 & 0 & \gamma _{2}^+ & 0 & 0 \\
			0 & 0 & 0 & 0 & 0 & -i g & 0 & 0 & 0 & -\frac{\Gamma }{2} & i g & 0 & 0 & 0 & 0 & 0 \\
			\gamma _{1}^- & 0 & 0 & 0 & 0 & 0 & -i g & 0 & 0 & i g & -\gamma _{1}^+ - \gamma _{2}^- & 0 & 0 & 0 & 0 & \gamma _{2}^+ \\
			0 & \gamma _{1}^- & 0 & 0 & 0 & 0 & 0 & -i g & 0 & 0 & 0 & -i \varepsilon -\gamma _{1}^+ -\frac{\Gamma _2}{2} & 0 & 0 & 0 & 0 \\
			0 & 0 & 0 & 0 & 0 & 0 & 0 & 0 & 0 & 0 & 0 & 0 & \frac{1}{2} (4 i \varepsilon -\Gamma ) & 0 & 0 & 0 \\
			0 & 0 & 0 & 0 & 0 & 0 & 0 & 0 & \gamma _{2}^- & 0 & 0 & 0 & 0 & i \varepsilon -\gamma _{2}^+ - \frac{\Gamma _1}{2} & i g & 0 \\
			0 & 0 & 0 & 0 & \gamma _{1}^- & 0 & 0 & 0 & 0 & 0 & 0 & 0 & 0 & i g & i \varepsilon -\gamma _{1}^+ -\frac{\Gamma _2}{2} & 0 \\
			0 & 0 & 0 & 0 & 0 & \gamma _{1}^-  & 0 & 0 & 0 & 0 & \gamma _{2}^- & 0 & 0 & 0 & 0 & -\gamma _{1}^+ -\gamma _{2}^+ \\
		\end{array}
		\right),$}
\end{equation}
where $\Gamma_k \coloneqq \gamma_k^+ +\gamma_k^-$ and $\Gamma \coloneqq \Gamma_1 +\Gamma_2$. A discussion on the spectrum of $L$ is given below.


\subsection{Spectrum}
\label{app:full_spectrum}

Except for $\lambda_1 = 0$,  all eigenvalues depend on the set of parameters $\xi = \{ \varepsilon,T_1,T_2, \gamma_1, \gamma_2,g\}$ and are shown in the complex plane in Fig.~\ref{fig:spectrum} a). The eigenvalues can be written in a neat form
\begin{equation}
	\begin{aligned}
		& \lambda_1 =0\,, \,  \lambda_{2} = -\Gamma \,,\lambda_{3,4} = -\frac{1}{2}\Gamma, \, \lambda_{5,6}=-\frac{1}{2}\Gamma\mp\eta \\ 
		& \lambda_{7,8} = \pm 2i\varepsilon - \frac{\Gamma}{2}\\
		& \lambda_{9,10} = \pm i\varepsilon -\frac{\Gamma}{2} - \sqrt{\alpha-\beta} \\
		& \lambda_{11,12} = \pm i\varepsilon -\frac{\Gamma}{2} + \sqrt{\alpha-\beta} \\
		& \lambda_{13,14} = \pm i\varepsilon - \frac{\Gamma}{2} - \sqrt{\alpha+\beta} \\
		& \lambda_{15,16} = \pm i\varepsilon -\frac{\Gamma}{2} + \sqrt{\alpha+\beta} \end{aligned}
\end{equation}
with the definitions $\eta \coloneqq \sqrt{\Delta \Gamma^2 - 4g^2}$, $\alpha \coloneqq -  g^2 + \frac{\Gamma_1^2 +\Gamma_2^2}{8}$, $\beta \coloneqq \frac{1}{8}\sqrt{4\left( \Delta \Gamma \cdot \Gamma\right)^2 - 16g^2 \delta}$, $\Delta \Gamma \coloneqq \left(\Gamma_1- \Gamma_2\right)/2$ and
\begin{equation}
	\begin{aligned}
		\delta \coloneqq \Gamma_1^2 + \Gamma_2^2 + 2\gamma_1^-\left(\gamma_2^- - 3\gamma_2^+\right) + 2\gamma_1^+\left(\gamma_2^+ - 3\gamma_2^-\right).
	\end{aligned}
\end{equation}

\begin{figure*}
	\hspace*{-1cm}
	\includegraphics[scale=0.4]{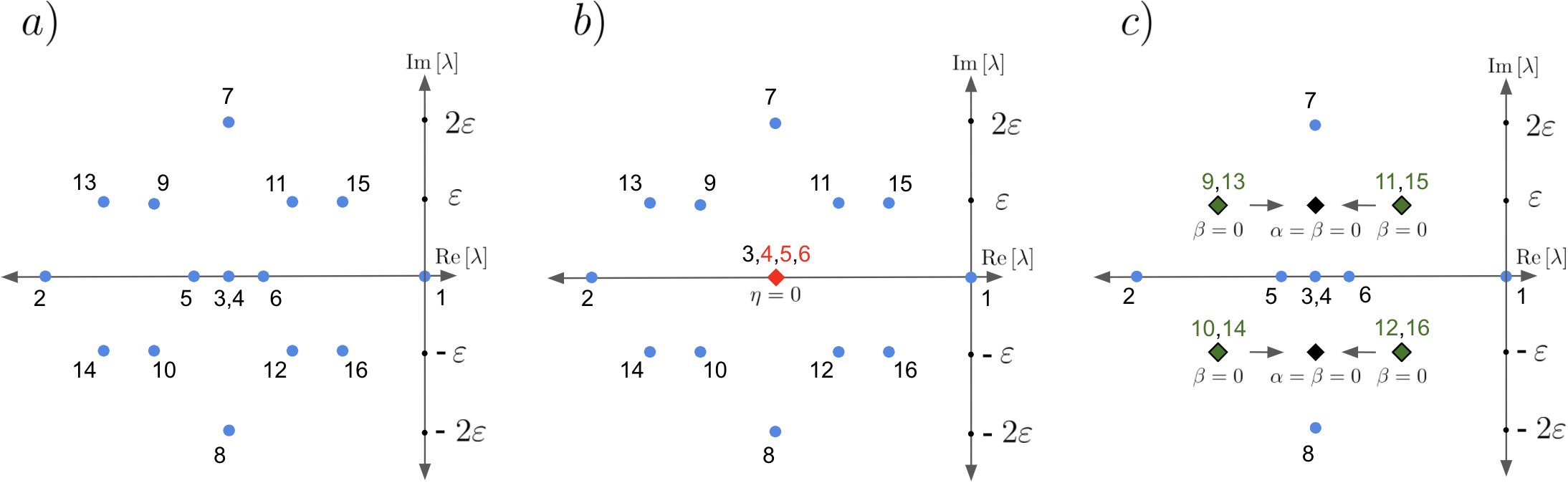}
	\caption{Representative eigenvalue configurations (specifically with $\eta\geq0$) for (a) a generic set of parameters, (b) the $\eta=0$ EP (in red), (c) simultaneous second-order EPs at $\beta=0$ (in green) turning into  simultaneous fourth-order EPs at $\alpha=\beta=0$ (in black).}\label{fig:spectrum}
\end{figure*}


 As expected, one of the eigenvalues of the Liouvillian is zero, and corresponds to the steady state \cite{Rivas12}. Furthermore, the eigenvalues are distributed symmetrically about the real axis owing to $L$ and $-L^*$ being isospectral \cite{Hatano19}. Although a tedious computation, the eigenvectors can be determined analytically using the above eigenvalues.

\subsection{Exceptional points}
\label{app:full_EP}

The exact form of all eigenvalues and eigenvectors allow us to list all possible EPs of the full Liouvillian:
\begin{enumerate}
	\item{$\eta = 0;\quad \lambda_4=\lambda_5=\lambda_6$ (third-order EP)}
	\item{$\beta=0;\quad\lambda_{9} = \lambda_{13},\,\lambda_{10} = \lambda_{14},\,\lambda_{11} = \lambda_{15}$ and $\lambda_{12} = \lambda_{16}$ (simultaneous second-order EPs)}
	\item{$\alpha-\beta=0;\quad \lambda_{9} = \lambda_{11},\,\lambda_{10} = \lambda_{12}$ (simultaneous second-order EPs)}
	\item{$\alpha+\beta=0; \quad \lambda_{13} = \lambda_{15},\,\lambda_{14} = \lambda_{16}$ (simultaneous second-order EPs)}
	\item{$\alpha=\beta=0; \quad\lambda_{9}=\lambda_{11}=\lambda_{13}=\lambda_{15},\, \lambda_{10}=\lambda_{12}=\lambda_{14}=\lambda_{16}$ (simultaneous fourth-order EPs)}
\end{enumerate}
Owing to fourth-degree polynomials appearing in the factorization of the characteristic polynomial, eight eigenvalues $\lambda_{9-16}$ are situated symmetrically on the complex plane. As a result, we can find sets of simultaneous second-order EPs (cases 2,3 and 4) and simultaneous fourth-order EPs (case 5). The simultaneous second-order EPs ($\beta=0$) turning into simultaneous fourth-order EPs ($\alpha =\beta=0$) are shown on the complex plane in Figure \ref{fig:spectrum} c). Note that there are no EPs in case of decoupled qubits (i.e. when $g=0$).\\

It must be noted that, in general, EPs are points in the parameter space that may not be necessarily reached for real values of the involved parameters (i.e. there may not be a real solution to the equations that are listed above). And although some EPs may be reached for real values of the parameters, they may or may not be reached for realistic (i.e. feasible) values of the parameters. For example the exceptional points discussed in \cite{Hatano19} can only be reached for strong coupling with the baths, in which case the validity of the master equation breaks down. In Fig.~\ref{fig:eps}, we plot the curves corresponding to $\eta=0$ and $\beta=0$ EPs in parameter space. Both types of EPs can be reached for a large range of reasonable parameters in the two-qubit thermal machine. We therefore show that this model is particularly suitable to investigate non-Hermitian physics in the context of quantum thermal machines directly amenable to experimental platforms.
\begin{figure}
	\centering
	\includegraphics[width=0.5\textwidth]{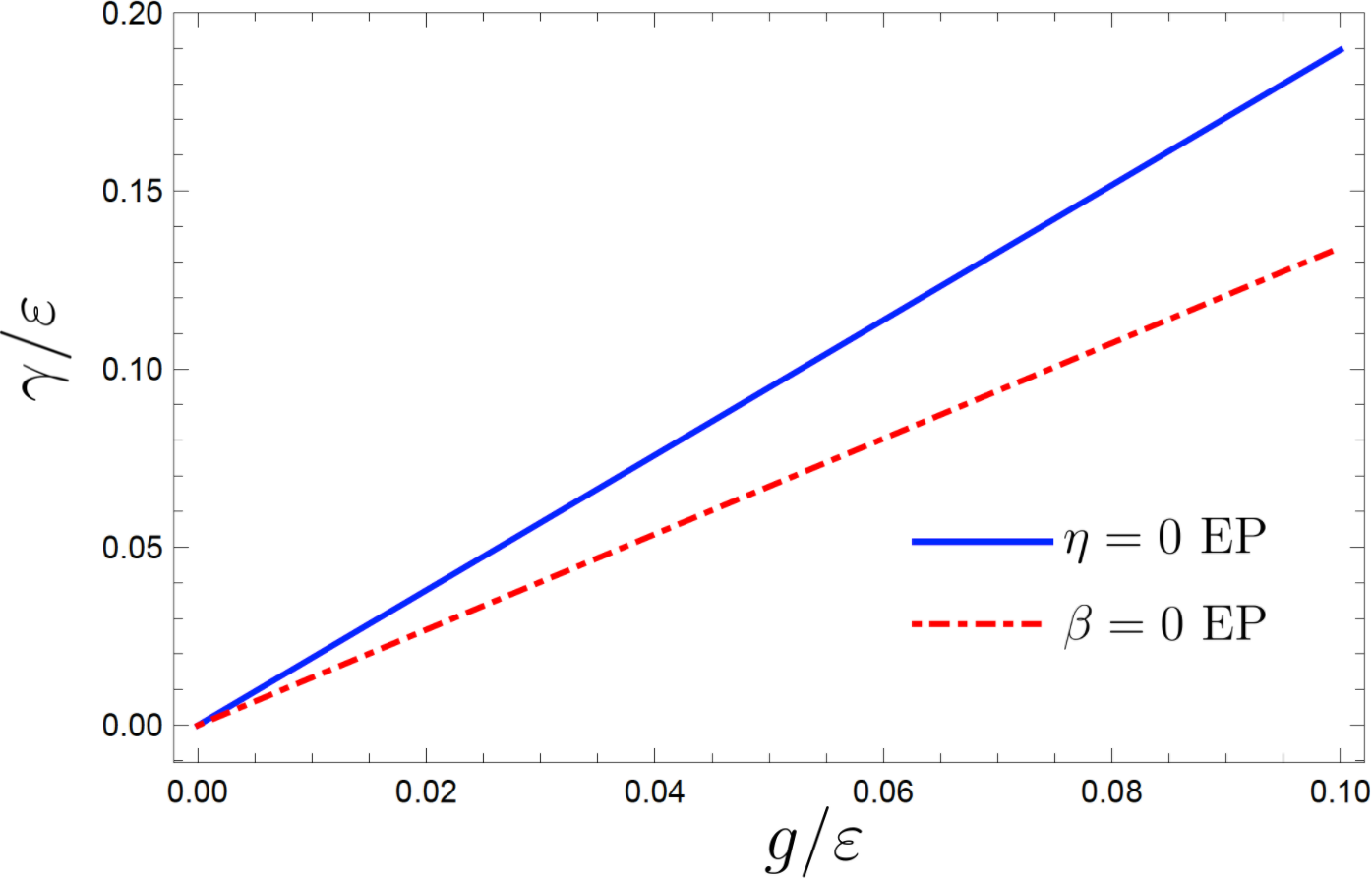}\caption{$\eta=0$ and $\beta=0$ exceptional points as curves in parameter space. $\gamma_1=\gamma_2=\gamma$, $T_1/\varepsilon=1.5$ and $T_2/\varepsilon=0.1$. The figure also shows the impossibility of reaching the two types of EPs simultaneously.}\label{fig:eps}
\end{figure}

\section{Effective Liouvillian $\tilde{L}$}
\label{app:effec}

\subsection{Compact form}
\label{app:compact}
As mentioned in the main text, if we restrict the initial state to have only the populations and coherences corresponding to $\ket{10}\bra{01}$ and $\ket{01}\bra{10}$, the $16\times16$ Liouvillian $L$ can be effectively reduced to a $6\times6$ Liouvillian $\tilde L$, given by

\begin{eqnarray}
 \left(\begin{array}{cccccc}
		-\gamma_1^{-}-\gamma_2^{-} & \gamma_2^{+} & \gamma_1^{+} & 0 & 0 & 0 \\
		\gamma_2^{-} & -\gamma_1^{-}-\gamma_2^{+} & 0 & \gamma_1^{+} & i g & -i g \\
		\gamma_1^{-} & 0 & -\gamma_1^{+}-\gamma_2^{-} & \gamma_2^{+} & -i g & i g \\
		0 & \gamma_1^{-} & \gamma_2^{-} & -\gamma_1^{+}-\gamma_2^{+} & 0 & 0 \\
		0 & i g & -i g & 0 & -\frac{\Gamma }{2} & 0 \\
		0 & -i g & i g & 0 & 0 & -\frac{\Gamma }{2} \\
	\end{array}
	\right).
\end{eqnarray}
The master equation for the reduced  $6\times1$ vectorized state $\tilde{\boldsymbol p}$ is then $\dot{\tilde{\boldsymbol p}} = \tilde L \tilde{\boldsymbol p}$. Due to the form of $\tilde{\boldsymbol p}$ (taking into account that the state has trace 1 and is Hermitian), the dimension of the Liouvillian can be further reduced by two. However, for completeness, we keep the effective Liouvillian $6\times6$. $\tilde L$ has the eigenvalues $\lambda_{1-6}$ of the full Liouvillian $L$. Assuming local detailed balance for the incoming and outgoing rates for each bath (independently of the bath statistics), $\gamma_k^+/\gamma_k^- = e^{-\varepsilon /T_k}$, $\tilde{L}$ can be written in a compact form reflecting the parameters' set of the model $\xi = \{\varepsilon, T_1, T_2, \gamma_1, \gamma_2,g\}$ with $f(T_k) = (1+ e^{\varepsilon/T_k})^{-1}$ the Fermi function of bath $k$ at temperature $T_k$,
	\begin{equation}
\tilde{L} =  \left( \!\begin{array}{cccccc}
	-\Gamma_1 f(-T_1)\!-\!\Gamma_2 f(-T_2) \! & \! \Gamma _{2} f(T_2)  & \Gamma _{1} f(T_1) & 0 & 0 & 0 \\
	\Gamma _{2} f(-T_2) & -\Gamma _{1} f(-T_1) -  \Gamma _{2} f(T_2)& 0 &\Gamma _{1} f(T_1)  & i g & -i g \\
	\Gamma _{1} f(-T_1) & 0 & - \Gamma _{1} f(T_1) -  \Gamma _{2} f(-T_2) &\Gamma _{2} f(T_2) & -i g & i g \\
	0 &\Gamma_1 f(-T_1) &\Gamma_2 f(-T_2) & -\Gamma _{1} f(T_1) -  \Gamma _{2} f(T_2)& 0 & 0 \\
	0 & i g & -i g & 0 & -\frac{\Gamma}{2} & 0 \\
	0 & -i g & i g & 0 & 0 & -\frac{\Gamma}{2} \end{array}\!\right)
\end{equation}
with $\Gamma_1 = \gamma_1^+ + \gamma_1^-, \Gamma_2 = \gamma_2^+ + \gamma_2^-$. For a fermionic bath, $\Gamma_k = \gamma_k$, whereas for a bosonic bath, $\Gamma_k = \gamma_k \tanh^{-1}(\varepsilon / (2 T_k))$. The sum $\Gamma = \Gamma_1 + \Gamma_2$ and difference $\Delta \Gamma = (\Gamma_1 - \Gamma_2)/2$, in interplay with the inter-qubit coupling $g$, will be the main parameters to reach EP in this reduced model. The spectrum $\sigma(\tilde{\mathcal{L}})$ reduces to $\{ \lambda_1 = 0,\,\, \lambda_2=-\Gamma,\,\,\lambda_{3,4}=-\frac{\Gamma}{2}\,, \lambda_{5,6}=-\frac{\Gamma}{2} \mp \eta \}$, with $\eta \coloneqq \sqrt{\Delta \Gamma^2 - 4g^2}$.\par
The class of initial states for which the results of  App. \ref{app:crit} remain unchanged, can be expanded by including coherences corresponding to $\ket{00}\bra{11}$ and $\ket{11}\bra{00}$. This means that we may consider arbitrary X-states as initial states. In such a case, a reduced $8\times 8$ Liouvillian is obtained with two additional eigenvalues $\lambda_{7,8}=\pm 2i\varepsilon - \Gamma/2$.

\subsection{Eigenmatrices}
\label{app:eigenmatrices}

Below, we provide the explicit expressions of the right eigenmatrices $\rho_{i}$ corresponding to the eigenvalues $\lambda_i$ of the above Liouvillian $\tilde{L}$. As mentioned in the text, the eigenmatrices can be constructed to be Hermitian and normalized according to the Hilbert-Schmidt inner product to ensure the unicity of the decomposition of $\rho(t)$, see Eq.~(2) in the main text.
\begin{equation}
	\hspace{-0.9cm}\rho_1=
	\resizebox{1.05\hsize}{!}{$ 
		\frac{1}{\Gamma^2 \left(4 g^2+\Gamma_1\Gamma_2\right)}\left(
		\begin{array}{cccc}
			4 g^2 (\gamma_1^{+}+\gamma_2^{+})^2+\gamma_1^{+} \gamma_2^{+} \Gamma^2 & 0 & 0 & 0 \\
			0 & 4 (\gamma_1^{-}+\gamma_2^{-}) (\gamma_1^{+}+\gamma_2^{+}) g^2+\gamma_1^{+} \gamma_2^{-} \Gamma^2 & 2 i g \Gamma(\gamma_1^{+} \gamma_2^{-}-\gamma_1^{-} \gamma_2^{+}) & 0 \\
			0 & -2 i g\Gamma (\gamma_1^{+} \gamma_2^{-} -\gamma_1^{-} \gamma_2^{+}) & 4 (\gamma_1^{-}+\gamma_2^{-}) (\gamma_1^{+}+\gamma_2^{+}) g^2+\gamma_1^{-} \gamma_2^{+} \Gamma^2 & 0 \\
			0 & 0 & 0 & 4 g^2 (\gamma_1^{-}+\gamma_2^{-})^2+\gamma_1^{-} \gamma_2^{-} \Gamma^2\\
		\end{array}
		\right) $}
\end{equation}
\begin{equation}
	\rho_2 = \frac{1}{2}\begin{pmatrix}
		1&0&0&0\\
		0&-1&0&0\\
		0&0&-1&0\\
		0&0&0&1
	\end{pmatrix} \quad\quad \rho_3 = \frac{1}{\sqrt{2}}\left(
	\begin{array}{cccc}
		0 & 0 & 0 & 0 \\
		0 & 0 &1& 0 \\
		0 & 1 & 0 & 0 \\
		0 & 0 & 0 & 0 \\
	\end{array}
	\right)
\end{equation}

\begin{equation}
\label{eq:eig4}
	\hspace*{-1cm}
	\rho_4 = \resizebox{1.05\hsize}{!}{$\frac{1}{N_4} \left(
		\begin{array}{cccc}
			\frac{2 (\gamma_1^{+}+\gamma_2^{+}) \sqrt{\Delta\Gamma^2-\eta ^2}}{ \Gamma\Delta\Gamma} & 0 & 0 & 0 \\
			0 & \frac{ (\gamma_1^{-}-\gamma_1^{+}+\gamma_2^{-}-\gamma_2^{+}) \sqrt{\Delta\Gamma^2-\eta ^2}}{\Gamma\Delta\Gamma} & -i & 0 \\
			0 & i & \frac{ (\gamma_1^{-}-\gamma_1^{+}+\gamma_2^{-}-\gamma_2^{+}) \sqrt{\Delta\Gamma^2-\eta ^2}}{\Gamma\Delta\Gamma} & 0 \\
			0 & 0 & 0 & -\frac{2 (\gamma_1^{-}+\gamma_2^{-}) \sqrt{\Delta\Gamma^2-\eta ^2}}{ \Gamma\Delta\Gamma} \\
		\end{array}
		\right)$}
\end{equation}
\begin{equation}
\label{eq:eig5}
	\hspace{-1.2cm}
	\resizebox{1.12\hsize}{!}{$
		\rho_{5,6} = \frac{1}{N_{5,6}}\left(
		\begin{array}{cccc}
			\frac{2\Delta\Gamma (\gamma_1^{+}+\gamma_2^{+}) \pm 2 (\gamma_1^{+}-\gamma_2^{+}) \eta }{(\Gamma+2 \eta ) \sqrt{\Delta\Gamma^2 -  \eta ^2}} & 0 & 0 & 0 \\
			0 & \frac{(\gamma_1^{-}-\gamma_1^{+}+\gamma_2^{-}-\gamma_2^{+})\Delta\Gamma-2 \eta ^2 \mp 2 (\gamma_1^{+}+\gamma_2^{-}) \eta }{(\Gamma + 2 \eta ) \sqrt{\Delta\Gamma^2 -  \eta ^2}} & -i & 0 \\
			0 & i & \frac{(\gamma_1^{-}-\gamma_1^{+}+\gamma_2^{-}-\gamma_2^{+})\Delta\Gamma+2 \eta ^2 \pm 2 (\gamma_1^{-}+\gamma_2^{+}) \eta }{(\Gamma+2 \eta ) \sqrt{\Delta\Gamma^2 -  \eta ^2}} & 0 \\
			0 & 0 & 0 & \frac{-2 \Delta\Gamma(\gamma_1^{-}+\gamma_2^{-}) \pm 2 (\gamma_2^{-}-\gamma_1^{-}) \eta }{(\Gamma+2 \eta ) \sqrt{\Delta\Gamma^2 -  \eta ^2}} \\
		\end{array}
		\right)$},
\end{equation}
where $\eta$, $\Gamma$, $\Gamma_{1,2}$ and $\Delta\Gamma$ were defined previously. $N_{4,5,6}$ are normalization factors for $\rho_{4,5,6}$ which we omit for brevity. Using Eqs. \eqref{eq:eig4} and \eqref{eq:eig5} can be checked that when $\eta=0$, $\rho_4,\rho_5$ and $\rho_6$ are identical - giving rise to a third-order EP. When $\eta=0$, $\lambda_3=\lambda_{4,5,6}$; however, the corresponding eigenmatrix $\rho_3$ clearly cannot coalesce with $\rho_{4,5,6}$. Therefore, the third eigenvalue does not have a role to play in the $\eta=0$ EP. \\ 


\subsection{Generalized eigenmatrices}
\label{app:gen}

At the $\eta=0$ EP, $\lambda_3=\lambda_4=\lambda_5=\lambda_6=\bar \lambda$ and $\rho_3\neq\rho_4=\rho_5=\rho_6$ (only the fourth, fifth and sixth eigenmatrices coalesce). The algebraic multiplicity of the eigenvalue $\bar\gamma$ is 4, while the geometric multiplicity is 2. Therefore, to form a complete basis of $\tilde L$ at the EP, it is necessary to find generalized eigenmatrices by constructing and solving a Jordan chain \cite{Horn2012}. It can be checked that the two following Jordan chains correspond to right and left generalized eigenmatrices. 

\begin{equation}
	\begin{aligned}
		&\left(\tilde{\mathcal{L}}- \bar{\lambda}\right)\rho_4= 0\\
		&\left(\tilde{\mathcal{L}}-\bar{\lambda}\right) \rho^{\prime}= \alpha\left(\bar\xi\right) \rho_4\\
		&\left(\tilde{\mathcal{L}}- \bar{\lambda}\right) \rho^{\prime\prime} = \beta\left(\bar\xi\right) \rho^{\prime},
	\end{aligned}
\end{equation}and 
\begin{equation}
	\begin{aligned}
		&\sigma_4^\dagger \left(\tilde{\mathcal{L}}- \bar{\lambda}\right) = \alpha\left(\bar\xi\right) \sigma^{\prime\dagger}\\
		&\sigma^{\prime\dagger}\left(\tilde{\mathcal{L}}-\bar{\lambda}\right) = \beta\left(\bar\xi\right) \sigma^{\prime\prime\dagger}\\
		&\sigma^{\prime\prime\dagger} \left(\tilde{\mathcal L} - \bar\lambda\right) =0.
	\end{aligned}
\end{equation}$\alpha$ and $\beta$ are in general arbitrary complex coefficients. However, if we demand normalization, as done in the main text, they depend on all parameters of the model.


\section{Strong inter-qubit coupling}
\label{app:strong}

In the case of strong inter-qubit coupling, one must rely on a \textit{global} master equation approach in which Lindblad jump operators are obtained by diagonalizing the system Hamiltonian. This case has been studied in detail in \cite{Hofer2017,Khandelwal2020}. The eigenenergies of $H_{\text{S}}+H_{\text{int}}$ are $0,\, \varepsilon_-=\varepsilon - g$, $\varepsilon_+=\varepsilon +g$ and $2\varepsilon$, and the corresponding eigenstates are denoted by $\ket{0}$, $\ket{\varepsilon_-}$, $\ket{\varepsilon_+}$ and $\ket{2}$, respectively. The only non-zero Lindblad operators are the ones corresponding to transitions of energies $\varepsilon_-$ and $\varepsilon_+$. For each of these energies, there are two possible transitions as $\varepsilon_-=\varepsilon_--0=2\varepsilon - \varepsilon_+$ and $\varepsilon_+=\varepsilon_+-0=2\varepsilon-\varepsilon_-$. The Lindblad operators are thus given by \cite{Hofer2017}
\begin{equation}
	\begin{aligned}
		&\hat L_j(\varepsilon_-) = \ket{0}\!\bra{0}\sigma_-^{\left(j\right)}\ket{\varepsilon_-}\!\bra{\varepsilon_-} + \ket{\varepsilon_+}\!\bra{\varepsilon_+}\sigma_-^{\left(j\right)}\ket{2}\!\bra{2}\\
		&\hat L_j(\varepsilon_+) = \ket{0}\!\bra{0}\sigma_-^{\left(j\right)}\ket{\varepsilon_+}\!\bra{\varepsilon_+} + \ket{\varepsilon_-}\!\bra{\varepsilon_-}\sigma_-^{\left(j\right)}\ket{2}\!\bra{2},
	\end{aligned}
\end{equation}and the adjoints of the above. The global master equation takes the form
\begin{equation}\label{eq:glind}
	\begin{aligned}
		\dot\rho (t)= -i\left[H_{\text{S}}+H_{\text{int}}, \rho(t)\right]\\
		+\sum_{\alpha\in\{-,+\}}\sum_{j\in\{h,c\}}&\gamma_j^{-}\left(\varepsilon_\alpha\right)\mathcal D\left[\hat L_j\left(\varepsilon_\alpha\right)\right]\rho(t)\\
		& +\gamma_j^{+}\left(\varepsilon_\alpha\right)\mathcal D\left[\hat L^\dagger_j\left(\varepsilon_\alpha\right)\right]\rho(t),
	\end{aligned}
\end{equation}where the dissipators $\mathcal D\left[A\right]$ were defined in the main text. The rates are determined by the eigenenergies $\varepsilon_-$ and $\varepsilon_+$,  $\gamma_j^-\left(\varepsilon_{\pm}\right)=\gamma_j\left(1+n_B^j\left(\varepsilon_{\pm}\right)\right)$ and $\gamma_j^+\left(\varepsilon_{\pm}\right)=\gamma_jn_B^j\left(\varepsilon_{\pm}\right)$ with $j=1,2$.\par
The $16\times16$ Liouvillian matrix can be constructed in the same way as for the weak inter-qubit coupling case. We list the 16 eigenvalues of this matrix below.
\begin{equation}
	\begin{aligned}
		&\lambda_1 = 0,\\& \lambda_2 = -\frac{1}{2}\left(\Gamma_1\left(\varepsilon_-\right) +\Gamma_2\left(\varepsilon_-\right)  \right), \quad \lambda_3 = -\frac{1}{2}\left(\Gamma_1\left(\varepsilon_+\right) +\Gamma_2\left(\varepsilon_+\right)\right),\\
		&\lambda_4 = -\frac{1}{2}\Gamma, \quad \lambda_{5,6} = \pm 2ig -\frac{1}{4}\Gamma, \quad \lambda_{7,8} = \pm2i\varepsilon -\frac{1}{4}\Gamma, \\
		&\lambda_{9,10} = \pm i\left(g+\varepsilon\right) -\frac{1}{4}\left(\Gamma+\sqrt{X}\right),\\
		&\lambda_{11,12} = \pm i\left(g+\varepsilon\right) -\frac{1}{4}\left(\Gamma-\sqrt{X}\right), \\
		&\lambda_{13,14} = \pm i\left(g-\varepsilon\right) -\frac{1}{4}\left(\Gamma+\sqrt{Y}\right), \\
		&\lambda_{15,16} = \pm i\left(g-\varepsilon\right) -\frac{1}{4}\left(\Gamma-\sqrt{Y}\right).
	\end{aligned}
\end{equation}with the definitions
\begin{equation}
	\begin{aligned}
		X\coloneqq &\Gamma_1\left(\varepsilon_-\right)^2+2 \gamma_1^-\left(\varepsilon_-\right) \left(\gamma_2^-\left(\varepsilon_-\right)-3 \gamma_2^+\left(\varepsilon_-\right)\right)\\&+2 \gamma_1^+\left(\varepsilon_-\right) \left(\gamma_2^+\left(\varepsilon_-\right)-3 \gamma_2^-\left(\varepsilon_-\right)\right)+\Gamma_2\left(\varepsilon_-\right)^2
	\end{aligned}
\end{equation}
\begin{equation}
	\begin{aligned}
		Y\coloneqq &\Gamma_1\left(\varepsilon_+\right)^2+2 \gamma_1^-\left(\varepsilon_+\right) \left(\gamma_2^-\left(\varepsilon_+\right)-3 \gamma_2^+\left(\varepsilon_+\right)\right)\\ &+2 \gamma_1^+\left(\varepsilon_+\right) \left(\gamma_2^+\left(\varepsilon_+\right)-3 \gamma_2^-\left(\varepsilon_+\right)\right)+\Gamma_2\left(\varepsilon_+\right)^2
	\end{aligned}
\end{equation}We find two sets of simultaneous second-order EPs:
\begin{enumerate}
	\item{$X=0$ $\implies$ $\lambda_9=\lambda_{11}$ and $\lambda_{10}=\lambda_{12}$.}
	\item{$Y=0$ $\implies$ $\lambda_{13}=\lambda_{15}$ and $\lambda_{14}=\lambda_{16}$\,,}
\end{enumerate}
for which we have checked the coalescence of their respective eigenvectors. Let us note that no EPs exist when considering the reduced Liouvillian (to the steady-state subspace), indicating that critical damping can not be achieved outside the weak interaction regime between the two qubits of the entanglement engine. See also Ref. \cite{Scali20} for a discussion of exceptional points as a function of the inter-qubit interaction strength.\\

\section{Proof of critical damping}
\label{app:crit}

To compare the long-time dynamics of the two-qubit system at the EP and at non-EPs (i.e. to compare how the system approaches the steady state in the two regimes), we use the trace distance defined in the main text, $ T\left(\rho,\sigma\right)\coloneqq \lvert\lvert  \rho-\sigma\rvert\rvert_1/2$. The solution to the master equation Eq.~\eqref{eq:lind} at non-EPs is given by
\begin{align}\label{eq:deco1}
	\rho(t, \xi) = \rho_{\text{ss}}(\xi) + \sum_{i=2}^{6} c_i(\xi) \, e^{\lambda_i(\xi) t}  \rho_i(\xi)\,.
\end{align}where $c_i\left(\xi\right) \coloneqq \langle \sigma_i\left(\xi\right),\rho\left(0\right) \rangle$ ($i=1,\cdot\cdot\cdot,6$) are scalars. While at the $\eta=0$ EP, the solution is given by (the ``bar" denoting the matrices/parameters at the EP)
\begin{equation}
	\begin{aligned}
		\bar \rho(t,\bar{\xi}) = &\bar\rho_{\text{ss}}(\bar{\xi}) +  \sum_{i=2,3} \bar{c}_i(\bar{\xi}) e^{\lambda_i (\bar{\xi}) t} \bar\rho_i(\bar{\xi}) \nonumber\\ & + \left( \bar{c}_4(\bar{\xi}) +\alpha(\bar{\xi}) \,  \bar{c}_5(\bar{\xi})  t +\alpha(\bar{\xi})\, \beta(\bar{\xi}) \,   \bar{c}_6(\bar{\xi})  \frac{t^2}{2} \right) e^{\bar{\lambda} t} \bar\rho_4(\bar{\xi}) \nonumber \\
		&+ \Big( \bar{c}_5(\bar{\xi})  + \beta(\bar{\xi}) \,  \bar{c}_6(\bar{\xi})  t \Big) e^{\bar{\lambda} t} \rho^{\prime}(\bar{\xi}) \nonumber +  \bar{c}_6(\bar{\xi})  e^{\bar{\lambda} t} \rho^{\prime\prime}(\bar{\xi})\,.
	\end{aligned}
\end{equation}where $\bar c_i\left(\bar\xi\right)\coloneqq \langle\bar\sigma_i\left(\bar\xi\right),\rho\left(0\right) \rangle$ $(i=1,\cdot\cdot\cdot,4)$, $\bar c_5\left(\bar\xi\right)\coloneqq \langle\sigma^\prime\left(\bar\xi\right),\rho\left(0\right) \rangle$ and $\bar c_6\left(\bar\xi\right)\coloneqq \langle\sigma^{\prime\prime}\left(\bar\xi\right),\rho\left(0\right) \rangle$. Considering the exact expressions of the eigenvalues of $\tilde{L}$, we exploit that
\begin{equation}
	\begin{aligned}
		& \lambda_2 = 2 \bar\lambda \\
		& \lambda_3 = \lambda_4 = \bar\lambda \\
		& \lambda_5 = \bar\lambda - \eta \\
		& \lambda_6 = \bar\lambda + \eta
	\end{aligned}
\end{equation}
to write the ratio of the trace distances to the steady state at the EP and at non-EPs as (for notational convenience, we omit $\xi$ and $\bar\xi$)
\begin{equation}\hspace*{-1.5cm}
	\begin{aligned}
		\mathcal R\left(t\right)\coloneqq \frac{T\left(\bar\rho(t),\bar\rho_{\text{ss}}\right)}{T\left(\rho(t),\rho_{\text{ss}}\right)} &=\frac{\left\lvert\left\lvert  \bar c_2e^{2\bar\lambda t} \bar\rho_2 + \bar c_3 e^{\bar\lambda t} \bar\rho_3  + \left(\bar c_4 + \alpha \bar c_5 t +\alpha\beta \bar c_6\frac{t^2}{2}\right)e^{\bar\lambda t} \bar\rho_4
			+\left(\bar c_5 + \beta \bar c_6t\right)e^{\bar\lambda t}\rho^{\prime} + \bar c_6e^{\bar\lambda t} \rho^{\prime\prime} \right\rvert\right\rvert_1}{\left\lvert\left\lvert  c_2e^{2\bar\lambda t} \rho_2  +c_3e^{\bar\lambda t} \rho_3 + c_4e^{\bar\lambda t} \rho_4 + c_5e^{\left(\bar\lambda - \eta\right) t} \rho_5 + c_6e^{\left(\bar\lambda +\eta\right) t} \rho_6 \right\rvert\right\rvert_1}\\
		&=\frac{\left\lvert\left\lvert  \bar c_2 e^{\bar\lambda t} \rho_2 + \bar c_3 \rho_3  + \left(\bar c_4 + \alpha \bar c_5 t +\alpha\beta \bar c_6\frac{t^2}{2}\right) \rho_4
			+\left(\bar c_5 + \beta \bar c_6t\right)\rho^{\prime} + \bar c_6 \rho^{\prime\prime} \right\rvert\right\rvert_1}{\left\lvert\left\lvert  c_2e^{\bar\lambda t} \rho_2 + c_3 \rho_3 + c_4 \rho_4 + c_5e^{-\eta t} \rho_5 + c_6e^{ +\eta t} \rho_6 \right\rvert\right\rvert_1}
	\end{aligned}
\end{equation}
If $\eta>0$, at long times, the terms proportional to $e^{\bar\lambda t}$ and $e^{-\eta t}$ vanish. The numerator therefore grows as $\mathcal O\left(t^2\right)$, while the denominator as $\mathcal O\left(e^{\eta t}\right)$. Therefore, at long times, we always have 
\begin{align}
	\frac{T\left(\bar \rho(t),\bar \rho_{\text{ss}}\right)}{T\left(\rho(t),\rho_{\text{ss}}\right)} <1, \quad \text{and} \quad \lim_{t\to\infty}\frac{T\left(\bar \rho(t),\rho_{\text{ss}}\right)}{T\left(\rho(t),\rho_{\text{ss}}\right)} =0.
\end{align}This shows that at long times, the system is closer to its steady state at the EP than at non-EPs, while preventing oscillatory dynamics, i.e. the approach to the steady state is the fastest aperiodic at the EP. The EP can therefore be seen as the point of critical damping. Note that for this result, it is essential that the initial state has some component in the sixth eigenmatrix (i.e. $c_6\coloneqq \langle\sigma_6,\rho\left(0\right) \rangle \neq 0$) as discussed in the main text.

\section{Analogy with the damped harmonic oscillator}
\label{app:ho}

It is well known that the exceptional point of a damped harmonic oscillator (HO) is the point of critical damping (see for example \cite{Fernandez2018}). In this section we provide details relevant for conducting the analogy between the two-qubit thermal machine we investigate in this work and the emblematic case of the damped HO. The equation of motion of a damped harmonic oscillator is given by
\begin{align}
	m\ddot x +2\gamma\dot x +k x=0.
\end{align}The regimes $\gamma^2<mk$, $\gamma^2>mk$ and $\gamma^2=mk$ determine whether the HO is underdamped (oscillatory), overdamped or critically damped, respectively. If we define $p=mx$, we can write the equation of motion in a vectorized form
\begin{align}\label{eq:harmonic}
	\frac{d}{dt}\begin{pmatrix}
		p\\
		x
	\end{pmatrix} = \begin{pmatrix}
		-2\gamma/m& -k\\
		1/m& 0
	\end{pmatrix}\begin{pmatrix}
		p\\
		x
	\end{pmatrix}\coloneqq M\begin{pmatrix}
		p\\
		x
	\end{pmatrix},
\end{align}where we refer to $M$ as the non-Hermitian evolution matrix. A simple calculation gives the eigenvalues of $M$, $\lambda_{1,2} = -\gamma/m\pm \sqrt{\gamma^2-mk}/m$ and the eigenvectors, $\boldsymbol v_{1,2} = \left(-\gamma/m\pm\sqrt{\gamma^2-mk}/m,1/m\right)^{T}$. Clearly, for $\gamma^2=mk$, the eigenvalues and eigenvectors both coalesce. Therefore, the system is critically damped at its EP and the time evolution matrix can be transformed to its Jordan canonical form. The transformation matrix $S$ can be constructed using the eigenvector and the generalized eigenvector $\boldsymbol v_2^{\text{EP}} \coloneqq (1,\gamma)^T$ \cite{Horn2012}.
\begin{align}
	S\coloneqq\begin{pmatrix}
		-\gamma& 1\\
		1& \gamma
	\end{pmatrix}=\left(\boldsymbol v_1^{\text{EP}} \quad \boldsymbol v_2^{\text{EP}} \right),\quad\quad S^{-1}\coloneqq \begin{pmatrix}
		\boldsymbol w_1^{\dagger\text{EP}}\\
		\boldsymbol w_2^{\dagger\text{EP}}
	\end{pmatrix},
\end{align}where $\boldsymbol w_j^{\text{EP}}$ are the left eigenvector and generalized eigenvector at the EP. The solution to Eq.~\eqref{eq:harmonic} is then given by
\begin{equation}
	\begin{aligned}
		\boldsymbol f_{\text{EP}}\left(t\right)&\coloneqq\begin{pmatrix}
			p(t)\\
			x(t)
		\end{pmatrix}=\\
	&= S\left(
		\begin{array}{cc}
			e^{-\frac{\gamma  t}{m}} & \frac{\left(\gamma ^2+1\right) e^{-\frac{\gamma  t}{m}} t}{m} \\
			0 & e^{-\frac{\gamma  t}{m}} \\
		\end{array}
		\right)S^{-1}\boldsymbol f_{\text{EP}}\left(0\right)\\
		&=\alpha^{\text{EP}}_1e^{\lambda t}\boldsymbol{v}_1^{\text{EP}} + \alpha^{\text{EP}}_2e^{\lambda t}\left(1+\left(\frac{\gamma^2+1}{m}\right)t\right)\boldsymbol{v}_2^{\text{EP}},
	\end{aligned}
\end{equation}where $\alpha^{\text{EP}}_j=\boldsymbol w_j^{\dagger\text{EP}}\boldsymbol f_{\text{EP}}\left(0\right)$ are scalars, and we have exponentiated the Jordan form in the middle. On the other hand, at non-EPs, the evolution matrix is diagonalizable and the solution is simply given by 
\begin{align}\label{eq:nonepdamp}
	\boldsymbol f\left(t\right) = \alpha_1e^{\lambda_1 t}\boldsymbol{v}_1 +\alpha_2e^{\lambda_2 t}\boldsymbol{v}_2,
\end{align}where $\alpha_j\coloneqq \boldsymbol w_j^{\dagger}\boldsymbol f\left(0\right)$, $\boldsymbol{w}_j$ being the left eigenvectors of $M$ and $\lambda \coloneqq-\gamma/m $.\par
In analogy to the ratio $\mathcal R(t)$ constructed for the two-qubit machine, we may construct $\mathcal R_{\text{HO}}(t)$ for the the damped harmonic oscillator, $\mathcal R_{\text{HO}}(t) \coloneqq d\left(\boldsymbol f_{\text{EP}}\left(t\right),\boldsymbol 0\right)/d\left(\boldsymbol f\left(t\right),\boldsymbol 0\right)$, where $d(\boldsymbol a, \boldsymbol b)\coloneqq \sqrt{\sum_j\left(a_j-b_j\right)^2}$ is the Euclidean distance between $\boldsymbol a$ and $\boldsymbol b$. $\mathcal R_{\text{HO}}<1$ implies that the state at the EP (i.e. at critical damping) is closer to relaxation than the state at non-EP, and $\mathcal R_{\text{HO}}>1$ implies the opposite. It is simple to show that at long times, we always have $\mathcal R_{\text{HO}}<1$ when critically damped; shown in Fig. \ref{fig:ratioHO}. It must be noted that this is true only when the initial state of the oscillator is not simply proportional to the second eigenvector (which would imply that $\alpha_1=0$). This is correspondence to what we find for the two-qubit machine, for which we had an analogous condition $c_6\neq 0$.

\begin{figure}
	\centering
	\includegraphics[width=0.5\textwidth]{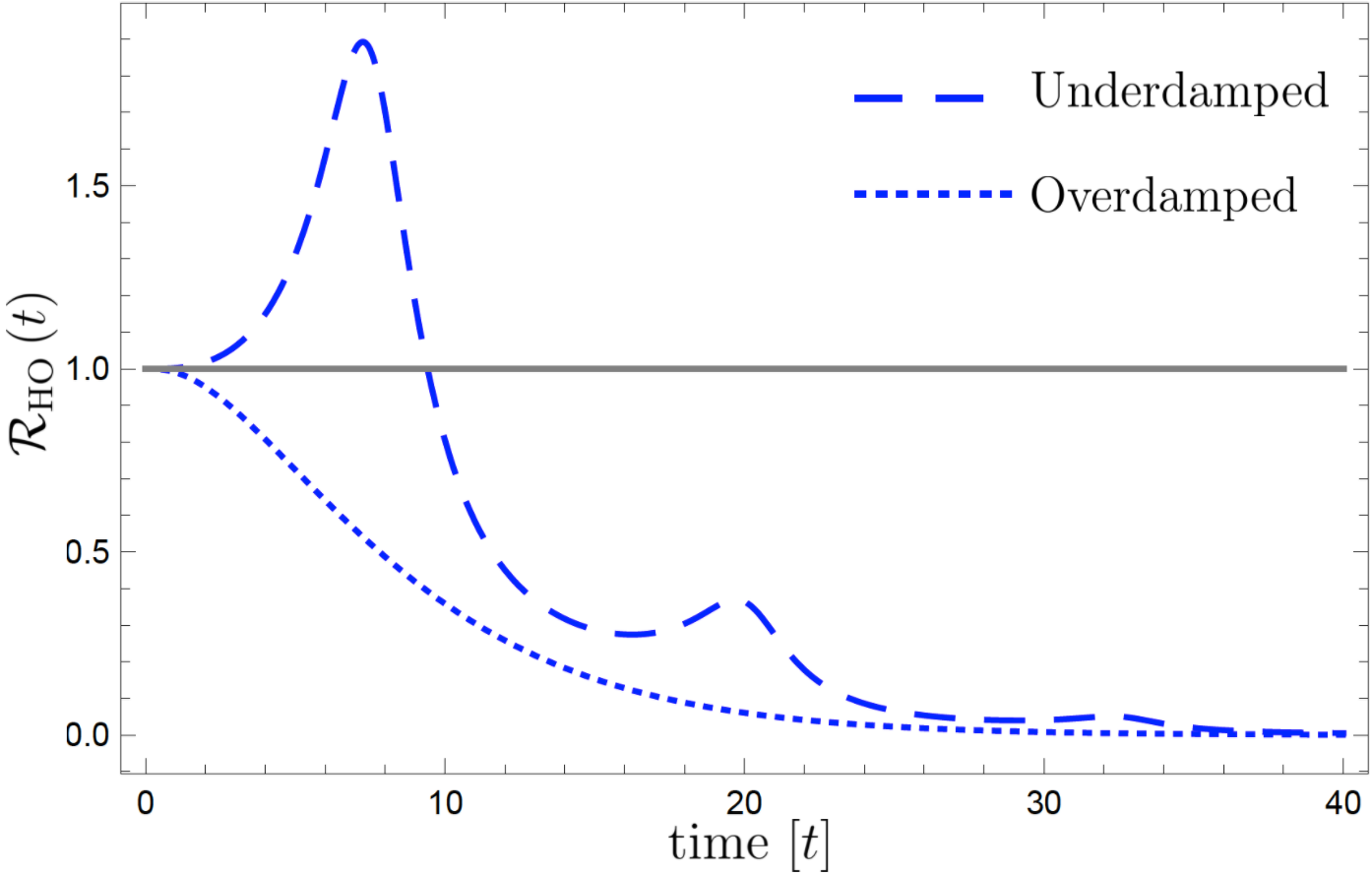}\caption{$\mathcal R_{\text{HO}}$ as a function of time. The initial state of the oscillator is an arbitrary sum of the two eigenvectors of $M$, the non-Hermitian evolution matrix of the damped HO. The ratio $\mathcal{R}$ is plotted considering the underdamped and overdamped regimes.}\label{fig:ratioHO}
\end{figure}

Similar to the two-qubit quantum thermal machine, if we take the initial state of the damped harmonic oscillator to be an eigenvector of the evolution matrix $M$, we can evidence a clear difference between the dynamics at the EP and non-EPs at short times. Indeed, taking $\boldsymbol{f}_{\text{EP}}\left(0\right) = a\boldsymbol v_2^{\text{EP}}$ $\left(a\in \mathbb R,\, \text{such that the initial state is a valid one}\right)$, we have   
\begin{align}\label{eq:epdamp}
	\boldsymbol f_{\text{EP}}\left(t\right) = ae^{\lambda t}\left(1+\left(\frac{\gamma^2+1}{m}\right)t\right)\boldsymbol{v}_2^{\text{EP}},
\end{align}which upon taking the logarithm will give linear and non-linear (logarithmic) terms in $t$. On the other hand at non-EPs, taking the initial state $\boldsymbol{f}\left(0\right) = a\boldsymbol{v}_2$ in Eq. \eqref{eq:nonepdamp}, we find $\boldsymbol f\left(t\right) = a e^{\lambda_2 t} \boldsymbol v_2$; which is linear on a logarithmic scale. The above equations are in complete analogy to the results obtained for the two-qubit thermal machine.

\end{widetext}

\begin{thebibliography}{0}
\expandafter\ifx\csname natexlab\endcsname\relax\def\natexlab#1{#1}\fi
\expandafter\ifx\csname bibnamefont\endcsname\relax
  \def\bibnamefont#1{#1}\fi
\expandafter\ifx\csname bibfnamefont\endcsname\relax
  \def\bibfnamefont#1{#1}\fi
\expandafter\ifx\csname citenamefont\endcsname\relax
  \def\citenamefont#1{#1}\fi
\expandafter\ifx\csname url\endcsname\relax
  \def\url#1{\texttt{#1}}\fi
\expandafter\ifx\csname urlprefix\endcsname\relax\def\urlprefix{URL }\fi
\providecommand{\bibinfo}[2]{#2}
\providecommand{\eprint}[2][]{\url{#2}}

\end{thebibliography}


\begin{thebibliography}{100}

\bibitem{Kosloff2014}R. Kosloff and A. Levy, \textit{Quantum heat engines and refrigerators: continuous devices. } \href{https://doi.org/10.1146/annurev-physchem-040513-103724}{Ann. Rev. Phys. Chem \textbf{65}, 365 (2014).}

\bibitem{Vinjanampathy2016}S. Vinjanampathy and J. Anders, \textit{Quantum thermodynamics},  \href{https://doi.org/10.1080/00107514.2016.1201896}{Contemp. Phys. \textbf{57}, 545 (2016).}


\bibitem{Mitchison}M. Mitchison, \textit{Quantum thermal absorption machines: refrigerators, engines and clocks,} \href{https://doi.org/10.1080/00107514.2019.1631555}{Contemp. Phys. \textbf{60}, 164 (2019).}

\bibitem{Bhattha20} S. Bhattacharjee and A. Dutta, \textit{Quantum thermal machines and batteries}, \href{}{arXiv:2008.07889 (2020).}



\bibitem{BohrBrask2015a}J.B. Brask, G. Haack, N. Brunner and M. Huber, \textit{Autonomous quantum thermal machine for generating steady-state entanglement} \href{https://iopscience.iop.org/article/10.1088/1367-2630/17/11/113029}{New J. Phys. \textbf{17}, 113029 (2015).} 

\bibitem{Tavakoli2018}A. Tavakoli, G. Haack, M. Huber and N. Brunner, \textit{Heralded generation of maximal entanglement in any dimension via incoherent coupling to thermal baths},  \href{https://doi.org/10.22331/q-2018-06-13-73}{Quantum \textbf{2}, 73 (2018).}

\bibitem{Khandelwal2020}S. Khandelwal, N. Palazzo, N. Brunner and G. Haack, \textit{Critical heat current for operating an entanglement engine,} \href{https://iopscience.iop.org/article/10.1088/1367-2630/ab9983}{New J. Phys. \textbf{22}, 073039 (2020).} 

\bibitem{Heineken2020} D. Heineken, K. Beyer, K. Luoma, W.T. Strunz, \textit{Quantum memory enhanced dissipative entanglement creation in non-equilibrium steady states}, \href{https://arxiv.org/abs/2008.04359}{arXiv:2008.04359 (2020).}


\bibitem{Breuer02} H. P. Breuer and F. Pettruccione, \textit{The Theory of Open Quantum Systems}, Oxford University Press (2007).	

\bibitem{Rivas12} A. Rivas, and S. F. Huelga, \textit{Open Quantum Systems}, Springer (2012).

\bibitem{Schaller14} G. Schaller, \textit{Open Quantum Systems Far from Equilibrium}, Springer (2014).

\bibitem{Kashuba13} O. Kashuba and H. Schoeller, \textit{Transient dynamics of open quantum systems}, \href{https://doi.org/10.1103/PhysRevB.87.201402}{Phys. Rev. B {\bf 87}, 201402(R) (2013).}

\bibitem{Hofer2017} P. P. Hofer, M. Perarnau-Llobet, L. D. M. Miranda, G. Haack, R. Silva, J. B. Brask and N. Brunner, \textit{Markovian master equations for quantum thermal machines: local versus global approach}, \href{https://iopscience.iop.org/article/10.1088/1367-2630/aa964f}{New J. Phys. \textbf{19}, 123037 (2017).}

\bibitem{Prosen08} T.Prosen, \textit{Third quantization: a general method to solve master equations for quadratic open Fermi systems}, \href{https://doi.org/10.1088/1367-2630/10/4/043026}{New J. Phys. {\textbf 10}, 043026 (2008).}

\bibitem{Brask15b} J. Bohr Brask and N. Brunner, \textit{Small quantum absorption refrigerator in the transient regime: Time scales, enhanced cooling, and entanglement},
\href{https://doi.org/10.1103/PhysRevE.92.062101}{Phys. Rev. E {\bf 92}, 062101 (2015). }

 \bibitem{Kshetrimayum17} A. Kshetrimayum, H. Weimer and R. Or\'us, \textit{A simple tensor network algorithm for two-dimensional steady states} \href{https://doi.org/10.1038/s41467-017-01511-6}{ Nat Commun {\bf 8}, 1291 (2017).}

\bibitem{Mitchison20} M. T. Mitchison, T. Fogarty, G. Guarnieri, S. Campbell, T. Busch, J. Goold, \textit{In Situ Thermometry of a Cold Fermi Gas via Dephasing Impurities}, \href{https://doi.org/10.1103/PhysRevLett.125.080402}{Phys. Rev. Lett. {\bf 125}, 080402 (2020).} 

\bibitem{Brenes20} M. Brenes, J. J. Mendoza-Arenas, A. Purkayastha, M. T. Mitchison, S. R. Clark, J. Goold, \textit{Tensor-Network Method to Simulate Strongly Interacting Quantum Thermal Machines}, \href{https://doi.org/10.1103/PhysRevX.10.031040}{Phys. Rev. X {\bf 10}, 031040 (2020).}

\bibitem{Haack20} G. Haack and A. Joye, \textit{Perturbation Analysis of Quantum Reset Models}, \href{https://doi.org/10.1007/s10955-021-02752-y}{J. Stat. Phys. \textbf{183}, 17 (2021).}

\bibitem{Goold20} A. Purkayastha, G. Guarnieri, S. Campbell, J. Prior, J. Goold, \textit{Periodically refreshed baths to simulate open quantum many-body dynamics}, \href{https://journals.aps.org/prb/abstract/10.1103/PhysRevB.104.045417}{Phys. Rev. B \textbf{104}, 045417 (2021).}

\bibitem{Ganainy18} R. El-Ganainy, K. Makris, M. Khajavikhan, Z. H. Musslimani, S. Rotter, N. D. Christodoulides, \textit{Non-Hermitian physics and PT symmetry}, \href{https://doi.org/10.1038/nphys4323}{Nat. Phys. {\bf14}, 11 (2018).}

\bibitem{Bender07} C. M. Bender, \textit{Making sense of non-Hermitian Hamiltonians}, \href{https://iopscience.iop.org/article/10.1088/0034-4885/70/6/R03/pdf}{Rep. Prog. Phys. {\bf 70}, 947 (2007).}

\bibitem{Rotter09} I. Rotter, \textit{A non-Hermitian Hamilton operator and the physics of open quantum systems}, \href{https://iopscience.iop.org/article/10.1088/1751-8113/42/15/153001/meta}{J. Phys. A: Math. Theor. {\bf 42}, 153001 (2009).}

\bibitem{Moiseyev11} N. Moiseyev, \textit{Non-Hermitian Quantum Mechanics}, Cambridge University Press (2011).

\bibitem{Minganti2018} F. Minganti, A. Biella, N. Bartolo, C. Ciuti, \textit{Spectral theory of Liouvillians for dissipative phase transitions}. \href{https://doi.org/10.1103/PhysRevA.98.042118}{Phys. Rev. A \textbf{98}, 042118 (2018).}


\bibitem{Heiss2012} W. D. Heiss, \textit{The physics of exceptional points,} \href{https://iopscience.iop.org/article/10.1088/1751-8113/45/44/444016}{J. Phys. A: Math. Theor. \textbf{45}, 444016 (2012).}

\bibitem{Minganti2019} F. Minganti, A. Miranowicz, R.W. Chhajlany, F. Nori, \textit{Quantum exceptional points of non-Hermitian Hamiltonians and Liouvillians: The effects of quantum jumps}.  \href{https://doi.org/10.1103/PhysRevA.100.062131}{Phys. Rev. A \textbf{100}, 062131 (2019).}

\bibitem{Miri19} M. Miri and A. Al\`u, \textit{Exceptional points in optics and photonics}, \href{https://science.sciencemag.org/content/363/6422/eaar7709}{Science \textbf{363}, 6422 (2019).}

\bibitem{Kato}T. Kato, \textit{Perturbation Theory for Linear Operators}, Springer (1995). 

\bibitem{Am15} M. Am-Shallem, R. Kosloff, N. Moiseyev, \textit{Exceptional points for parameter estimation in open quantum systems: analysis of the Bloch equations}, \href{https://iopscience.iop.org/article/10.1088/1367-2630/17/11/113036/pdf}{New J. Phys. {\bf 17}, 113036 (2015).}

\bibitem{Chen2017} W. Chen, \c{S}.K. \"{O}zdemir, G. Zhao, J. Wiersig, L. Yang, \textit{Exceptional points enhance sensing in an optical microcavity,} \href{https://doi.org/10.1038/nature23281}{Nature \textbf{548}, 192 (2017).}

\bibitem{Hodaei2017} H. Hodaei, A.U. Hassan, S. Wittek, H. Garcia-Gracia, \textit{Enhanced sensitivity at higher-order exceptional points
}, \href{https://doi.org/10.1038/nature23280}{Nature \textbf{548}, 187 (2017).}

\bibitem{Lau18} H.K. Lau and A.A. Clerk, \textit{Fundamental limits and non-reciprocal approaches in non-Hermitian quantum sensing}, \href{https://doi.org/10.1038/s41467-018-06477-7}{Nat. Comm. {\bf 9}, 4320 (2018).}

\bibitem{H2019} C. Chen, L. Jin and R.B. Liu, \textit{Sensitivity of parameter estimation near the exceptional point of a non-Hermitian system}, \href{https://iopscience.iop.org/article/10.1088/1367-2630/ab32ab}{New J. Phys. \textbf{21} 083002 (2019).}

\bibitem{Djorwe2019} P. Djorwe, Y. Pennec, and B. Djafari-Rouhani, \textit{Exceptional Point Enhances Sensitivity of Optomechanical Mass Sensors}, \href{https://doi.org/10.1103/PhysRevApplied.12.024002}{Phys. Rev. Applied \textbf{12}, 024002 (2019).}



\bibitem{Peng14} B. Peng, \c{S}.K. \"Ozdemir, S. Rotter, H. Yilmaz, M. Liertzer, F. Monifi, C. M. Bender, F. Nori, L. Yang, \textit{Loss-induced suppression and revival of lasing}, \href{https://science.sciencemag.org/content/346/6207/328}{Science \textbf{346}, 328 (2014).}

\bibitem{Gao15} T. Gao, E. Estrecho, K. Y. Bliokh, T. C. H. Liew et al., \textit{Observation of non-Hermitian degeneracies in a chaotic exciton-polariton billiard}. \href{https://doi.org/10.1038/nature15522}{Nature \textbf{526}, 554 (2015).} 

\bibitem{Ozdemir19}\c{S}.K. \"Ozdemir, S. Rotter, F. Nori and L. Yang, \textit{Parity--time symmetry and exceptional points in photonics,} \href{ https://doi.org/10.1038/s41563-019-0304-9}{Nat. Mat. \textbf{18}, 783 (2019).}




\bibitem{Partanen18} M. Partanen, J. Goetz, K.Y. Tan, K. Kohvakka et al. \textit{Exceptional points in tunable superconducting resonators}. \href{https://doi.org/10.1103/PhysRevB.100.134505}{Phys. Rev. B \textbf{100}, 134505 (2019).}
\bibitem{Naghiloo19} M. Naghiloo, M. Abbasi, Y.N. Joglekar and K.W. Murch. \textit{Quantum state tomography across the exceptional point in a single dissipative qubit}. \href{https://doi.org/10.1038/s41567-019-0652-z}{Nat. Phys. \textbf{15}, 1232 (2019).}



\bibitem{Hatano19} N. Hatano, \textit{Exceptional points of the Lindblad operator of a two-level system,} \href{https://doi.org/10.1080/00268976.2019.1593535}{Molecular Phys. \textbf{117}, 2121 (2019).}




\bibitem{Dietz07} B. Dietz, T. Friedrich, J. Metz, M. Miski-Oglu, A. Richter,F. Sch\"{a}fer, and C. A. Stafford, \textit{Rabi oscillations at exceptional points in microwave billiards}, \href{https://doi.org/10.1103/PhysRevE.75.027201}{Phys. Rev. E {\bf 75}, 027201 (2007).}

\bibitem{Cartarius11} H. Cartarius and N. Moiseyev, \textit{Fingerprints of exceptional points in the survival probability of resonances in atomic spectra}, \href{https://doi.org/10.1103/PhysRevA.84.013419}{Phys. Rev. A {\bf84}, 013419 (2011).}

\bibitem{Demange12} G. Demange and E.-M. Graefe, \textit{Signatures of three coalescing eigenfunctions}, \href{https://iopscience.iop.org/article/10.1088/1751-8113/45/2/025303/meta}{J. Phys. A: Math. Theor. {\bf 45}, 025303 (2012).}

\bibitem{Uzdin2013} R. Uzdin, E.G.D. Torre, R. Kosloff, and N. Moiseyev, \textit{Effects of an exceptional point on the dynamics of a single particle in a time-dependent harmonic trap}, \href{http://dx.doi.org/10.1103/PhysRevA.88.022505}{Phys. Rev. A \textbf{88}, 022505 (2013).}

\bibitem{Gilary13} I. Gilary, A. A. Mailybaev, and N. Moiseyev, \textit{Time-asymmetric quantum-state-exchange mechanism}, \href{https://doi.org/10.1103/PhysRevA.88.010102}{Phys. Rev. A {\bf88}, 010102(R) (2013).}

\bibitem{Garmon17} S. Garmon and G. Ordonez, \textit{Characteristic dynamics near two coalescing eigenvalues incorporating continuum threshold effects}, \href{https://doi.org/10.1063/1.4983809}{J. of Math. Phys. {\bf58}, 062101 (2017).}

\bibitem{Kosloff2017}R. Kosloff and Y. Rezek, \textit{The Quantum Harmonic Otto Cycle}, \href{https://doi.org/10.3390/e19040136}{Entropy \textbf{19}, 136 (2017).} 

\bibitem{Insinga18} A. Insinga, B. Andresen, P. Salamon, R. Kosloff, \textit{Quantum heat engines: limit cycles and exceptional points}, \href{https://doi.org/10.1103/PhysRevE.97.062153}{Phys. Rev. E {\bf 97}, 062153 (2018).}

\bibitem{Fernando18} F. Quijandria, U. Naether, S. K. \"{O}zdemir, F. Nori, and D. Zueco, \textit{PT-symmetric circuit QED}, Phys. Rev. A \textbf{97}, 053846 (2018).

\bibitem{Archak20} A. Purkayastha, M. Kulkarni, and Y. N. Joglekar, \textit{Emergent PT symmetry in a double-quantum-dot circuit QED setup}, \href{https://doi.org/10.1103/PhysRevResearch.2.043075}{Phys. Rev. Research {\bf 2}, 043075 (2020).}

\bibitem{Chakraborty19} S. Chakraborty and A. K. Sarma, \textit{Delayed sudden death of entanglement at exceptional points}, \href{https://doi.org/10.1103/PhysRevA.100.063846}{Phys. Rev. A {\textbf 100}, 063846 (2019).} 

\bibitem{Cabot19} A. Cabot, G. L. Giorgi, S. Longhi and R. Zambrini, \textit{Exceptional points in 1D arrays of quantum harmonic oscillators}, \href{https://iopscience.iop.org/article/10.1209/0295-5075/127/20001}{Euro. Phys. Lett. \bf{127}, 20001 (2019).}










\bibitem{Macieszczak2016} K. Macieszczak, M. Gu\c{t}\v{a}, I. Lesanovsky, J.P. Garrahan, \textit{Towards a Theory of Metastability in Open Quantum Dynamics}, \href{https://doi.org/10.1103/PhysRevLett.116.240404}{Phys. Rev. Lett. \textbf{116}, 240404 (2016).}

\bibitem{Ding16} K. Ding, G. Ma, M. Xiao, Z. Q. Zhang, and C. T. Chan, \textit{Emergence, Coalescence, and Topological Properties of Multiple Exceptional Points
	and Their Experimental Realization}, \href{https://doi.org/10.1103/PhysRevX.6.021007}{Phys. Rev. X {\bf 6}, 021007 (2016).}

\bibitem{Bhattacherjee19} S. Bhattacherjee, H. K. Gandhi, A. Laha, and S. Ghosh, \textit{Higher-order topological degeneracies and progress towards unique successive state switching in a four-level open system}, \href{https://doi.org/10.1103/PhysRevA.100.062124}{Phys. Rev. A {\bf 100}, 062124 (2019).} 

\bibitem{Dey20} S. Dey, A. Laha, S. Ghosh, \textit{Exotic light dynamics around an exceptional point of order four associated with three connecting second-order exceptional points}, \href{https://doi.org/10.1364/JOSAB.416232}{ J. Opt. Soc. Am. B \textbf{38}, 1297-1306 (2021).}




\bibitem{Horn2012} R. A. Horn and C. R. Johnson, \textit{Matrix Analysis}, Cambridge University Press (2012).

\bibitem{Fernandez2018} F.M. Fernandez, \textit{Exceptional point in a simple textbook example}, \href{https://iopscience.iop.org/article/10.1088/1361-6404/aab6df/meta}{Eur. J. Phys. \textbf{39}, 045005 (2018).} 

\bibitem{Fernandez2020}L.J. Fern\'andez-Alc\'azar, R. Kononchuk, T. Kottos, \textit{Enhanced energy harvesting near exceptional points in systems with (pseudo-)PT-symmetry}, \href{https://doi.org/10.1038/s42005-021-00577-5}{Comm. Phys. \textbf{4}, 79 (2021.}

\bibitem{Kumar2021} P. Kumar, H.G. Zirnstein, K. Snizhko, Y. Gefen and B. Rosenow, \textit{Optimized Quantum Steering and Exceptional Points}, \href{https://arxiv.org/abs/2101.07284}{arXiv:2101.07284.}

\bibitem{Scali20} S. Scali, J. Anders, L. A. Correa, \textit{Local master equations bypass the secular approximation}, \href{https://doi.org/10.22331/q-2021-05-01-451}{Quantum {\bf5}, 451 (2021).}


\end{thebibliography}

\end{document}